\documentclass[aps,prc,superscriptaddress,showpacs,amssymb,amsmath,amsfonts,floatfix]{revtex4}
\usepackage{graphicx}
\usepackage{color}

\begin{document}

%%%%%%%%%%%%%%%%%%%%%%%%%%%%%%%%%%%%%%%%%%%%%%%%%%
\title{Sensitivity of the COSY Dibaryon Candidate
to $np$ Elastic Scattering Measurements}

%%%%%%%%%%%%%%%%%%%%%%%%%%%%%%%%%%%%%%%%%%%%%%%%%%%%
%%%%%%%%%%%%%%%%%%%%%%%%%%%%%%%%%%%%%%%%%%%%%%%%%%%%%%%%%%%%%%%%%%%%%
\newcommand*{\GWU}{Institute for Nuclear Studies, Department of Physics, 
            The George Washington University, Washington, DC 20052, USA}
\affiliation{\GWU}

\author {R.~L.~Workman}
\affiliation{\GWU}
\author {W.~J.~Briscoe}
\affiliation{\GWU}
\author {I.~I.~Strakovsky}
\affiliation{\GWU}
%%%%%%%%%%%%%%%%%%%%%%%%%%%%%%%%%%%%%%%%%%%%%%%%%%%%%%%%%%%%%%%%%%%%

%%%%%%%%%%%%%%%%%%%%%%%%%%%%%%%%%%%%%%%%%%%%%%%%%%%%%
\begin{abstract}
The case for a dibaryon resonance, appearing in $np$ scattering,
has support from a WASA-at-COSY measurement of the polarization
quantity $A_y$ over a center-of-mass energy region suggested by
structures seen earlier in two-pion production experiments.
Here we compare fits with and without an associated pole in order
to clarify the impact of these COSY data. We then consider what
further $np$ scattering measurements would most clearly distinguish 
between the pole and non-pole fit results.
\end{abstract}

%%%%%%%%%%%%%%%%%%%%%%%%%%%%%%%%%%%%%%%%%%%%%%%%%%%%%

\pacs{11.80.Et,13.75.Cs,25.40.Cm,25.40.Dn}
\maketitle

%%%%%%%%%%%%%%%%%%%%%%%%%%%%%%%%%%%%%%%%%%%%%%%%%%%%%
\section{Introduction}
\label{sec:intro}

The longitudinal spin-dependent proton-proton total cross-section
difference $\Delta\sigma_L$ measurements at the zero-gradient 
synchrotron (ZGS)~\cite{Auer} stimulated a high level of experimental 
and theoretical activity to search for dibaryons, mostly with 
isospin~1, through the 80's.  Details of this period can be found
in reviews~\cite{Yokosawa,Locher,IIS}. In the end, the difficulties 
in distinguishing true and pseudo-resonances~\cite{pseudo}
led to the demise of these investigations. A post-mortem is
given in Ref.~\cite{seth}.

\begin{figure}[th]
\centerline{
\includegraphics[height=0.35\textwidth, angle=90]{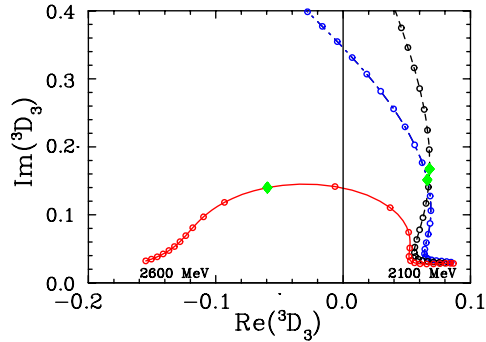}}
\vspace{3mm}
\caption{(Color online) Argand plot for the dimensionless $^3D_3$ 
	$np$ amplitude.
	Previous SAID SP07 solution shown as a black dashed 
	line~\protect\cite{sp07}. Revised SAID solution without 
	(with) a pole is plotted as a blue dot-dashed (red solid) 
	line. Energies are plotted with open circles in 20-MeV 
	steps.  Green filled diamond symbols correspond to the 
	pole mass $W_R = 2380~MeV$. \label{fig:g1}}
\end{figure}

In a recent series of WASA-at-COSY two-pion production 
measurements~\cite{wasa_2pi}, a resonance-like structure was 
reported, corresponding to an isospin~0 
resonance mass near 2.38~GeV, with a width of about 70~MeV. This 
claim gained added weight with the analysis of $A_y$ data from 
$np$ elastic scattering, also measured by the WASA-at-COSY 
Collaboration, which showed a rapid variation centered near the 
2.38~GeV CM energy~\cite{wasa_prl,wasa_prc}. The most recent GW 
SAID~\cite{sp07} $NN$ partial wave analysis (PWA) was not able 
to predict this behavior, nor was it present in previous 
fits~\cite{said_NN}. However, a re-analysis of the full database, 
including the COSY measurements, 
resulted in the generation of a pole. The location of the pole, 
seen in the coupled $^3D_3$-$^3G_3$ partial waves, was 
[(2380$\pm$10) - $i(40\pm 5)$]~MeV, 
corresponding almost exactly to the earlier resonance mass and 
width estimates~\cite{wasa_prl,wasa_prc,krakow}. An associated
Argand plot of the $^3D_3$ partial-wave is shown in Fig.~\ref{fig:g1}.  

The close 
correspondence of this resonance energy with a very early 
prediction, within the SU(6) quark model
of Dyson and Huong~\cite{dyson}, is also remarkable. Given the 
resurgence of interest in states beyond the usual $q\bar{q}$ 
mesons and $qqq$ baryons, there have been numerous publications 
focused on related states and strategies for their 
detection~\cite{brodsky,bashkanov15}. With this motivation, we 
have made a more detailed study of the structure found through 
the analysis of $np$ elastic scattering data. In particular, we 
consider those additional measurements which are most sensitive 
to the pole structure.
%%%%%%%%%%%%%%%%%%%%%%%%%%%%%%%%%%%%%%%%%%%%%%%%%%%%%%
%%%%%%%%%%%%%%%%%%%%%%%%%%%%%%%%%%%%%%%%%%%%%%%%%%%%%%

%%%%%%%%%%%%%%%%%%%%%%%%%%%%%%%%%%%%%%%%%%%%%%%%%%%%%%%%%%
\section{Fits to the COSY $A_y$ data}
\label{sec:AY_Data}

\begin{figure}[th]
\centerline{
\includegraphics[height=0.35\textwidth, angle=90]{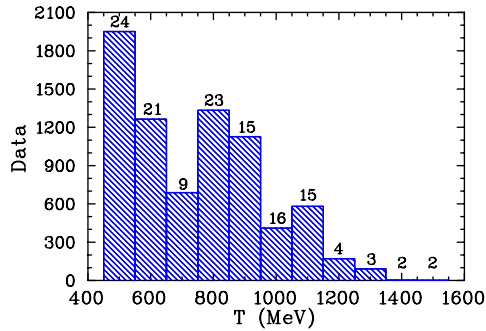}}
\vspace{3mm}
\caption{(Color online) Data available for $np\to np$ as 
	a function of neutron kinetic 
	energy~\protect\cite{SAID}. The number of observable 
	types is given above each vertical bar. 
	\label{fig:g2}}
\end{figure}

The COSY experiment measured 7 angular distributions for 
$A_y$ in $np$ scattering using a polarized deuteron beam 
impinging on a hydrogen target~\cite{wasa_prl}. The resulting 
neutron kinetic energies ranged from 1.108 to 1.197~GeV, 
corresponding to CM energies between 2.367 and 2.403~GeV. As 
shown in Fig.~\ref{fig:g2} the number of existing data, and data 
types, drops of rapidly beyond a kinetic energy of 1.1~GeV and 
this limits the reliability of PWA much beyond 1.1~GeV. The 
GW SAID PWA~\cite{sp07} has an upper limit of 1.3~GeV for an 
energy-dependent fit and the highest energy for an $np$ 
amplitude reconstruction is 1.1~GeV in Ref.~\cite{ball}.
%%%%%%%%%%%%%%%%%%%%%%%%%%%%%%%%%%%%%%%%%%%%%%%%%%%%%%
%%%%%%%%%%%%%%%%%%%%%%%%%%%%%%%%%%%%%%%%%%%%%%%%%%%%%%

\begin{figure*}[th]
\centerline{
\includegraphics[height=0.35\textwidth, angle=90]{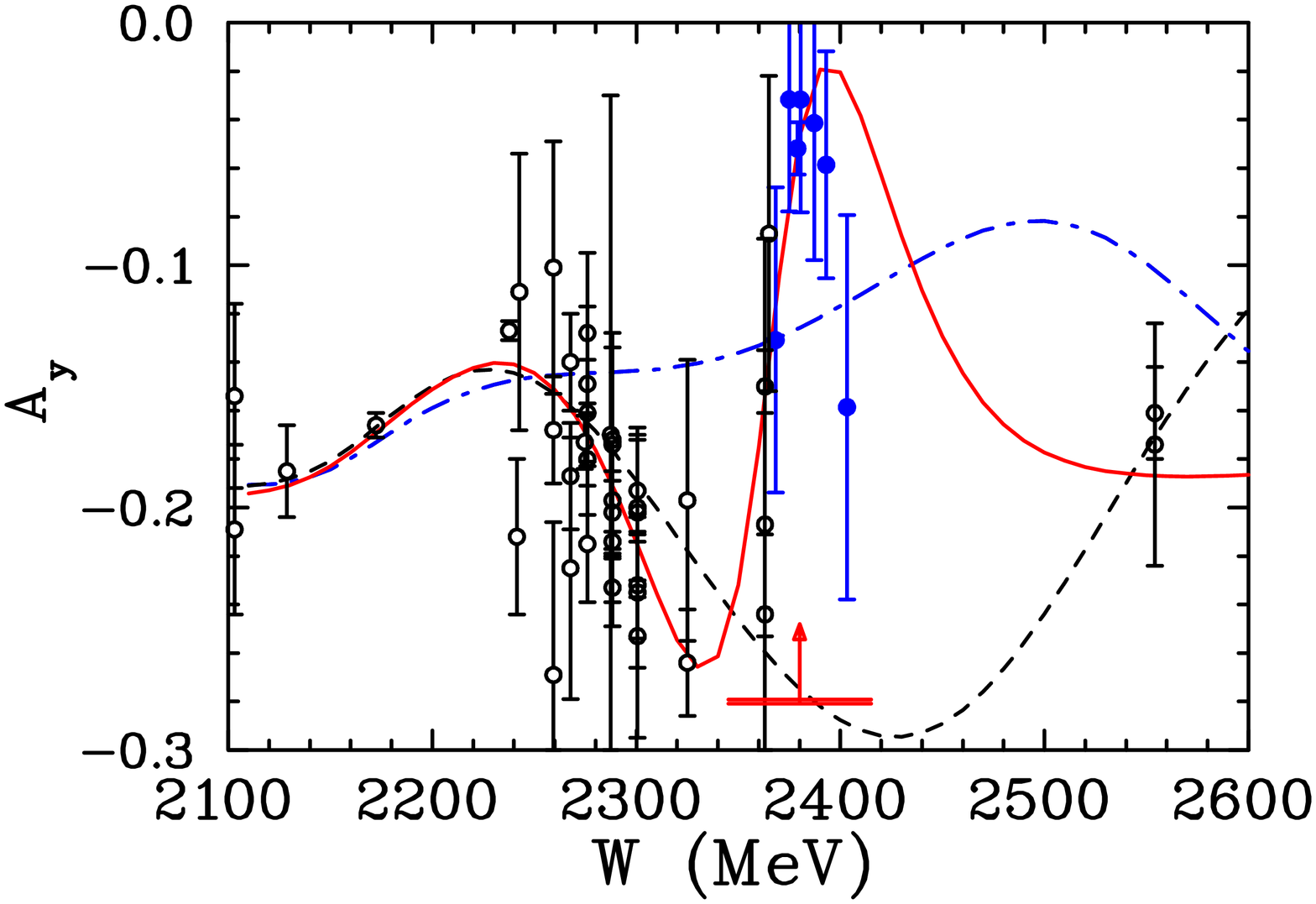}\hfill
\includegraphics[height=0.35\textwidth, angle=90]{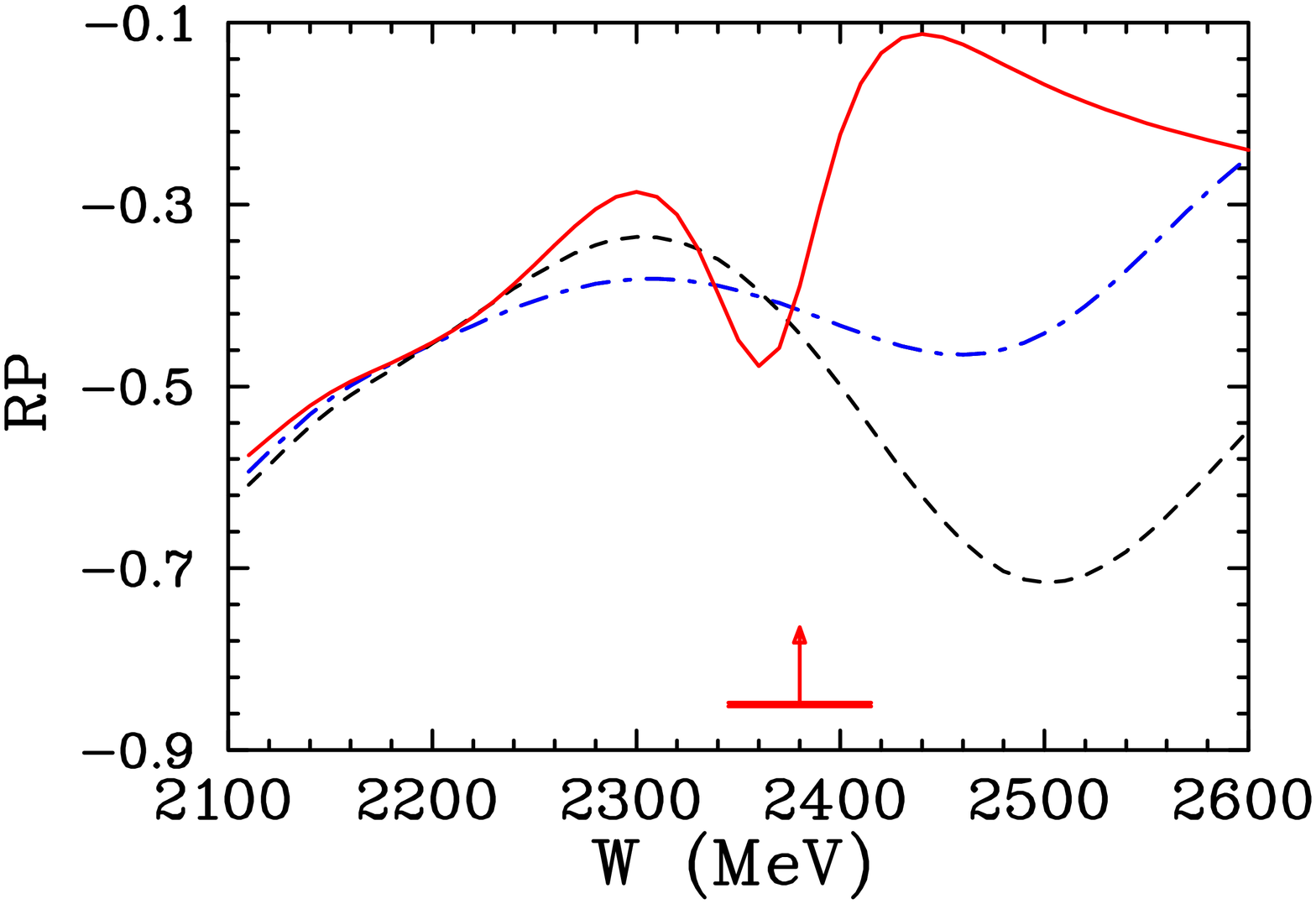}\hfill
\includegraphics[height=0.35\textwidth, angle=90]{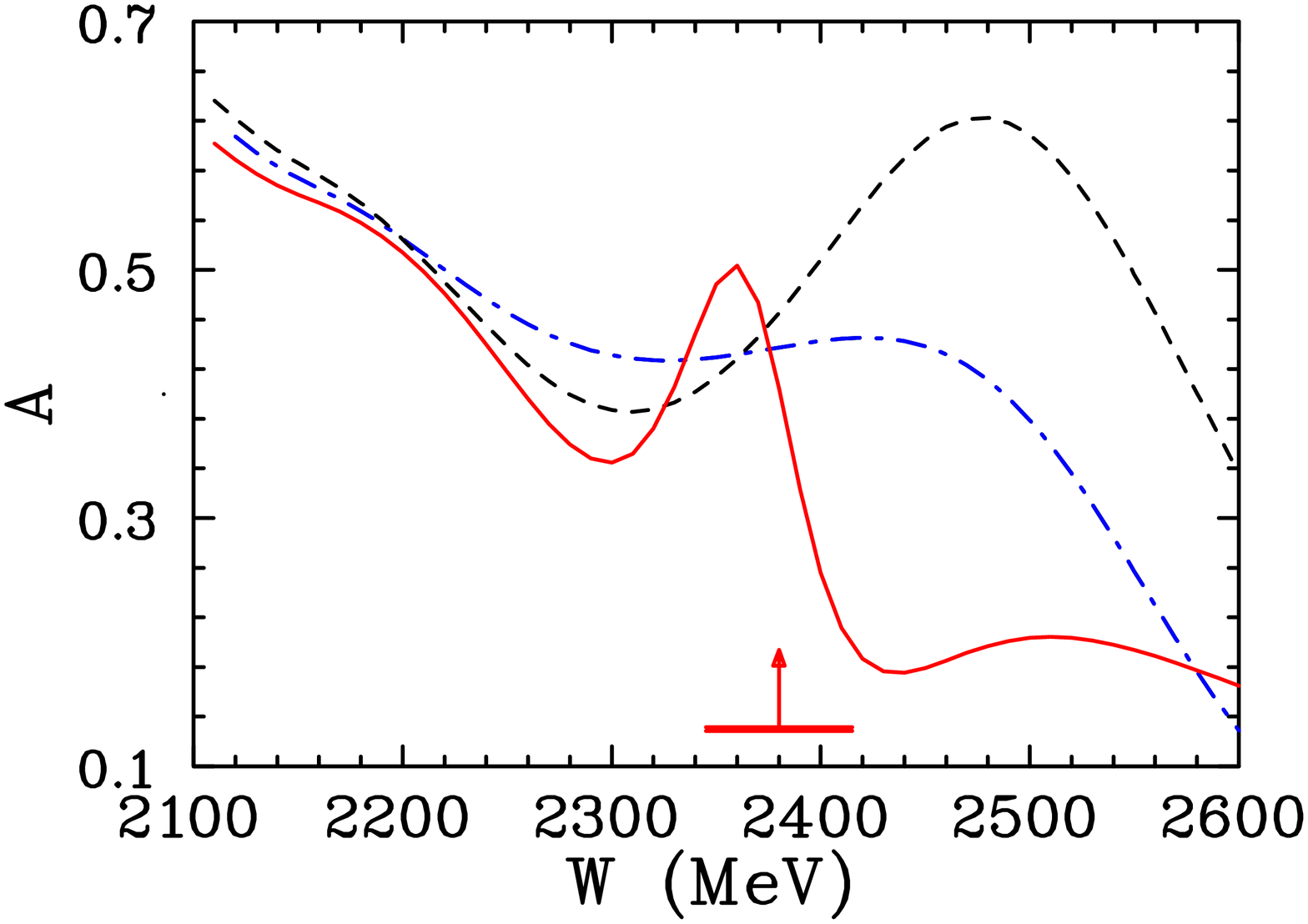}}
\centerline{
\includegraphics[height=0.35\textwidth, angle=90]{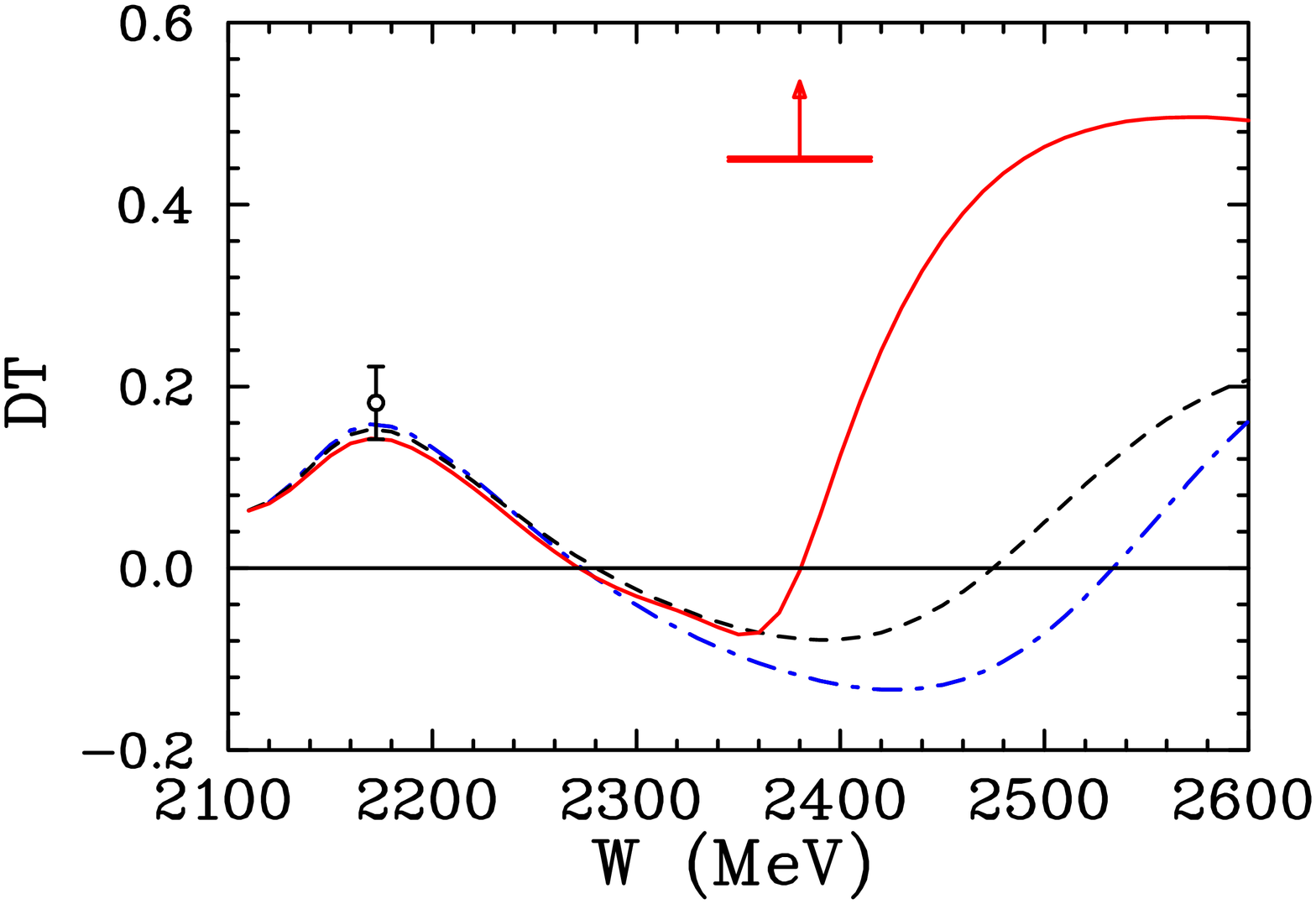}\hfill
\includegraphics[height=0.35\textwidth, angle=90]{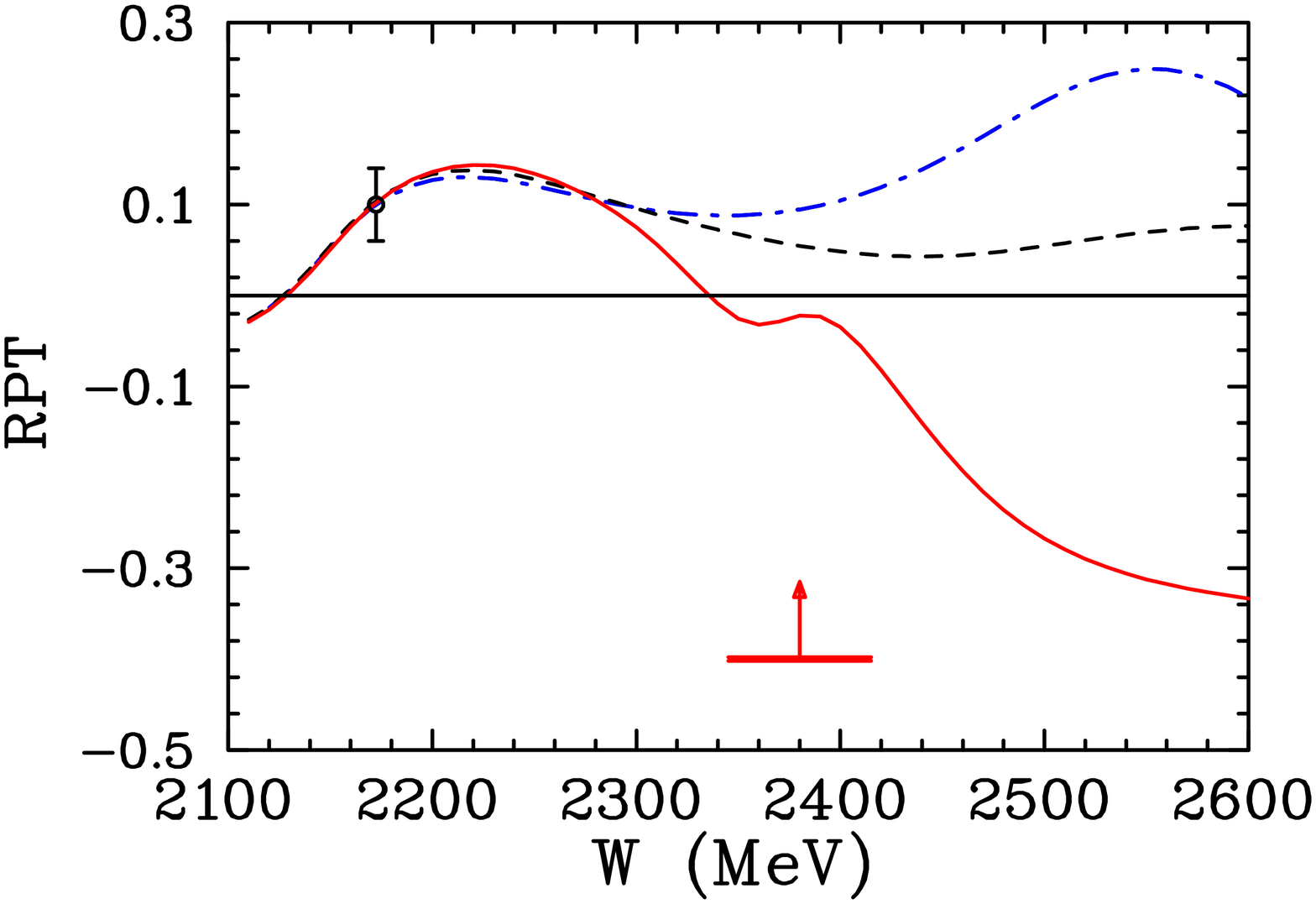}\hfill
\includegraphics[height=0.35\textwidth, angle=90]{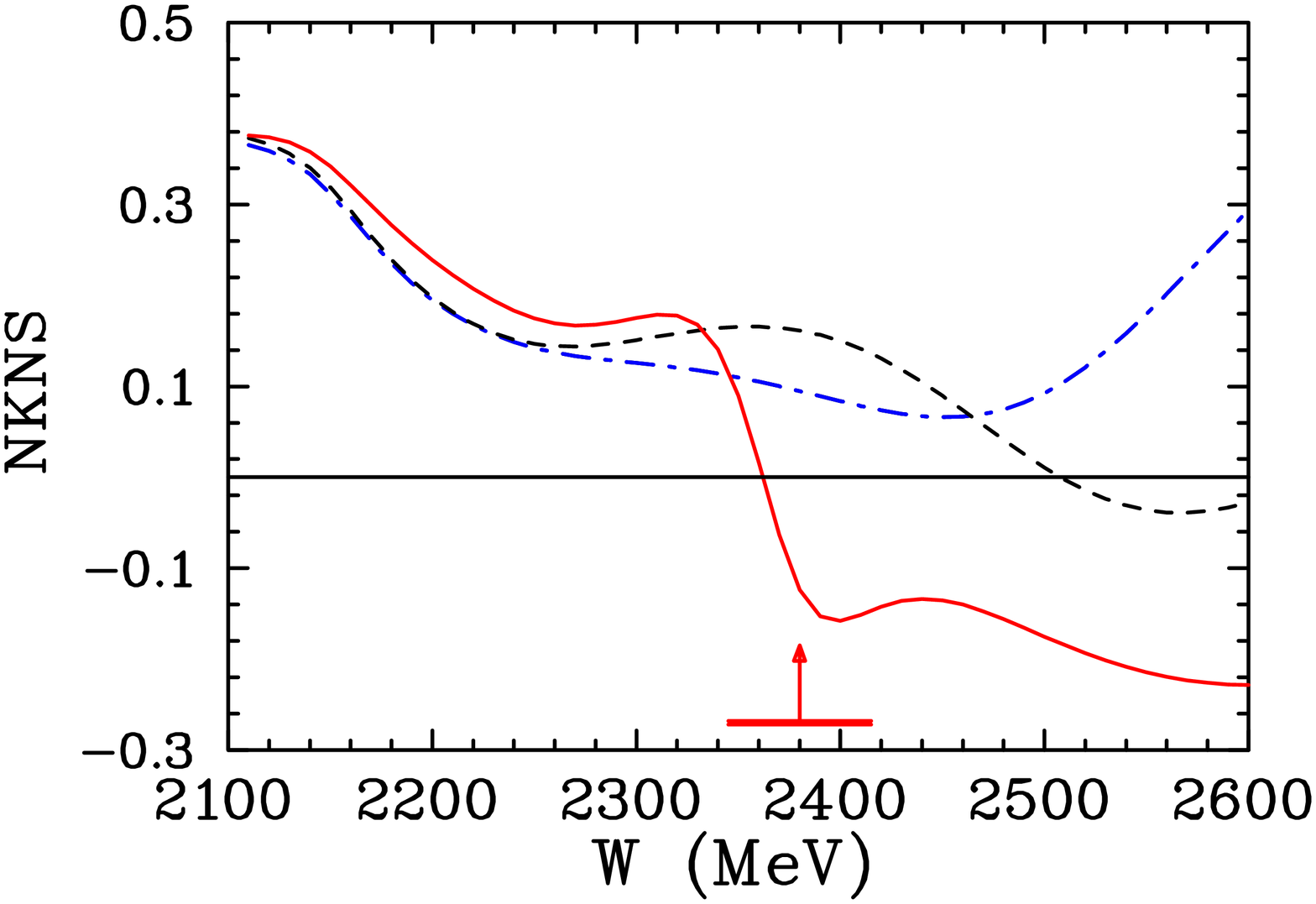}}
\centerline{
\includegraphics[height=0.35\textwidth, angle=90]{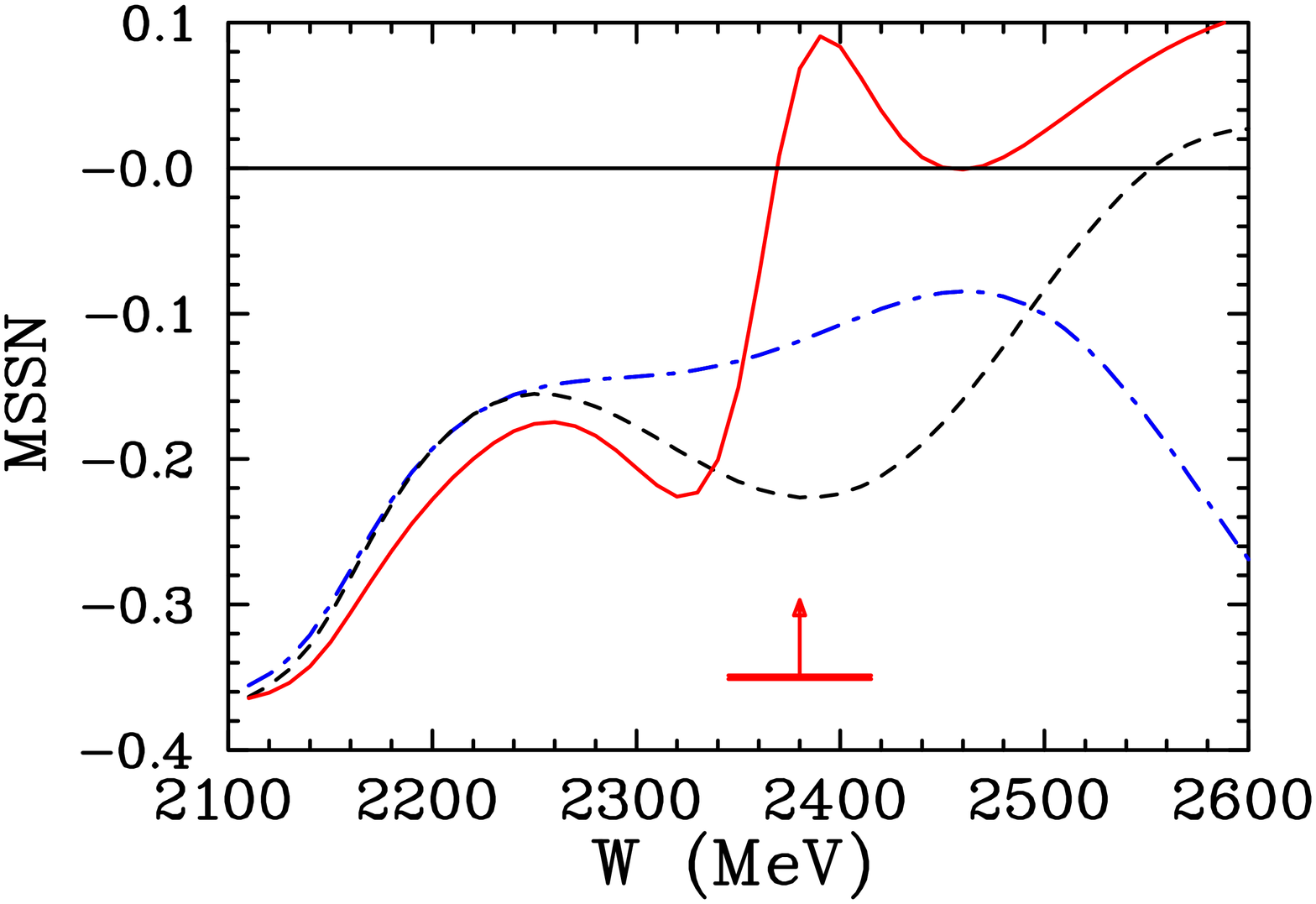}\hfill
\includegraphics[height=0.35\textwidth, angle=90]{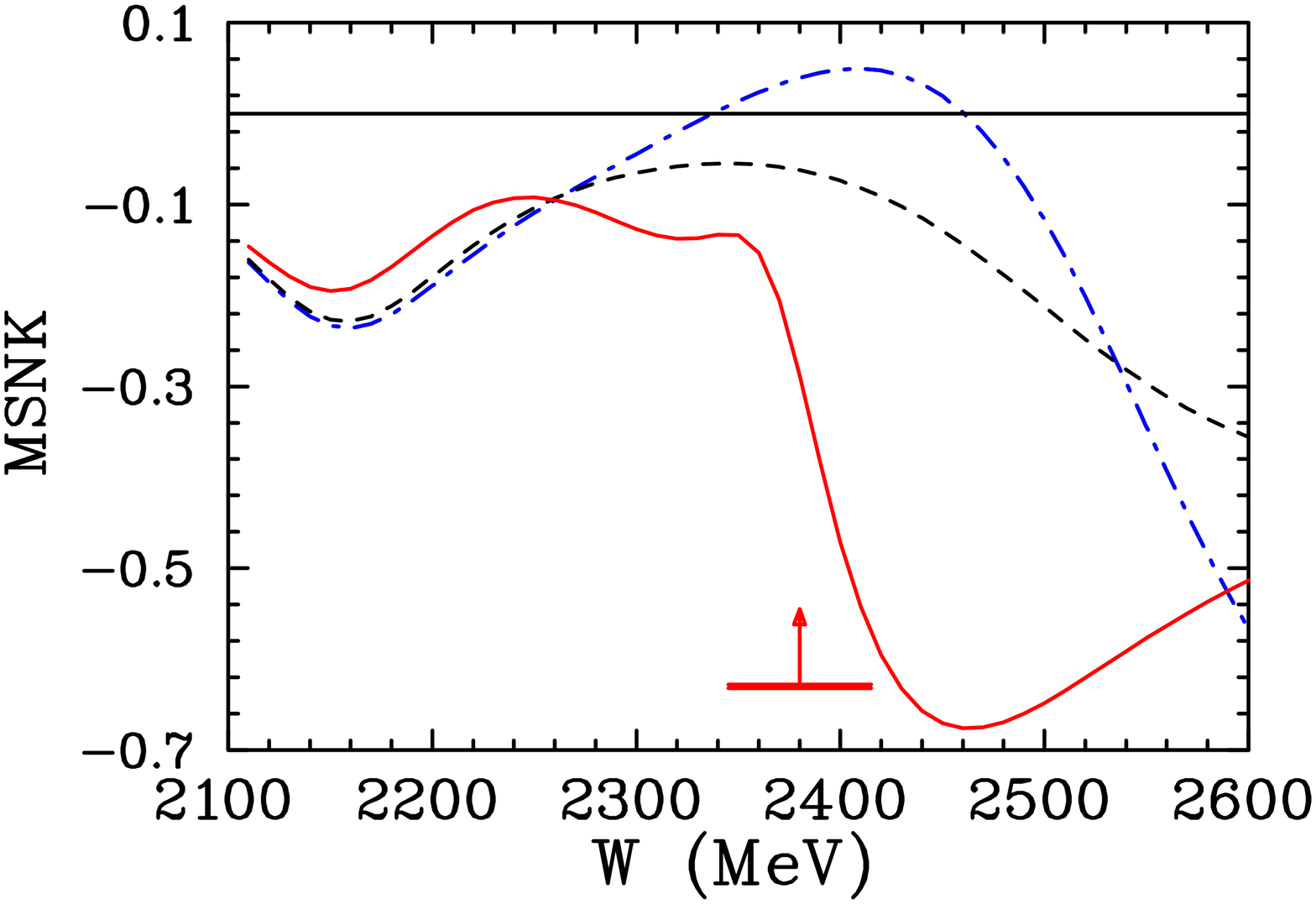}\hfill
\includegraphics[height=0.35\textwidth, angle=90]{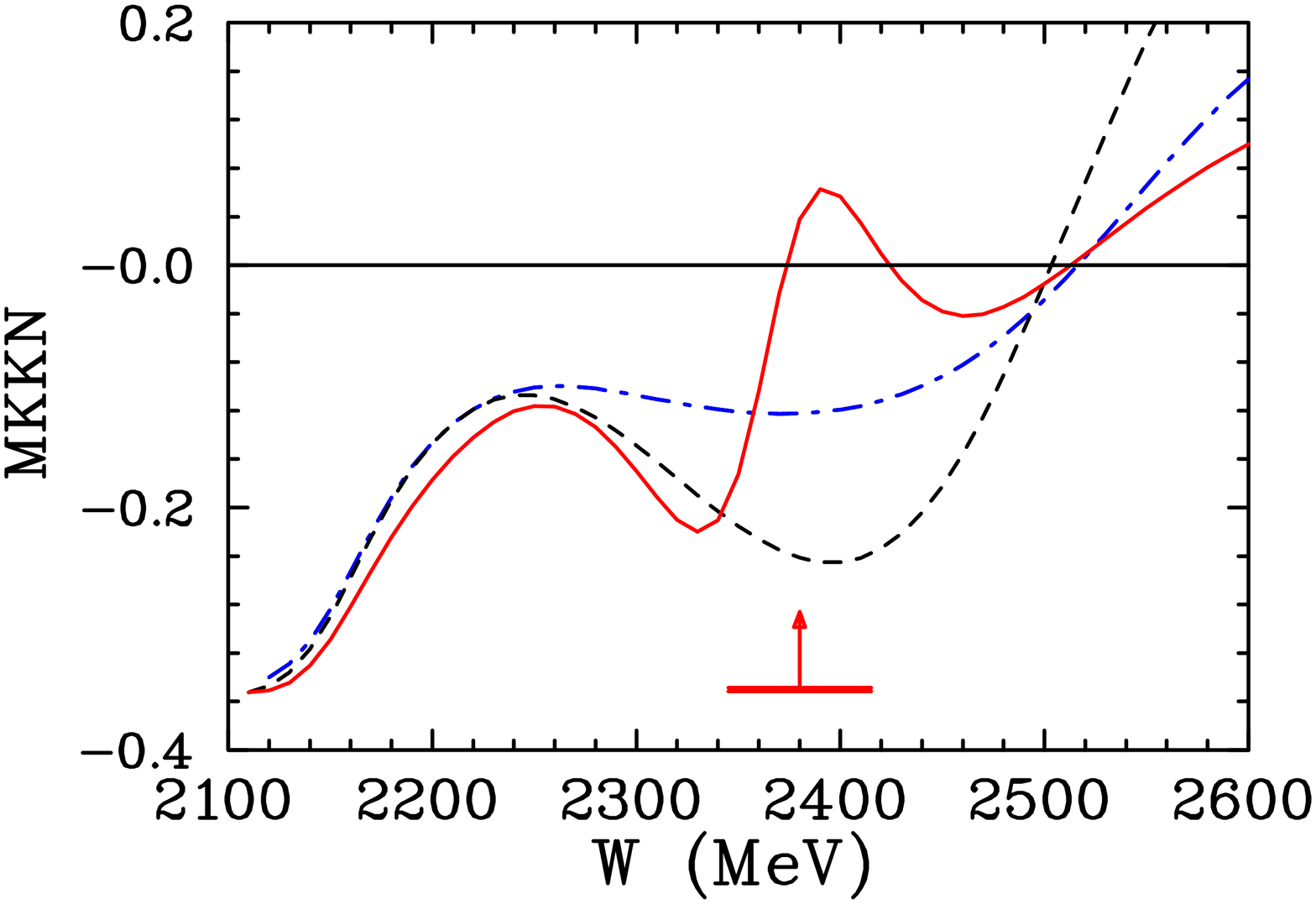}}
\vspace{3mm}
\caption{(Color online) Polarized observables for energies near
        the COSY resonance~\protect\cite{wasa_prl}, at 
        $85^\circ$. Data shown as blue solid circles are from the
        COSY experiment~\protect\cite{wasa_prl}. Previous
        measurements within $\Delta\theta =
	\pm 1^\circ$~\protect\cite{SAID} shown as black open 
	circles. SAID SP07 solution shown as a black dashed 
	line~\protect\cite{sp07}. Revised fit without (with) a 
	pole displayed 
        by blue dot-dashed (red solid) line. Red vertical 
	arrows indicate resonance mass $W_R$ value and red 
	horizontal bar gives the full width 
	$\Gamma$~\protect\cite{wasa_prl}. \label{fig:g3}}
\end{figure*}

In Fig.~\ref{fig:g3}, the SP07 prediction~\cite{sp07} and the 
fit containing a pole are compared to data at 85$^\circ$ -- 
where the variation is greatest. The SP07 prediction clearly 
misses the rapid rise in $A_y$ (top left) while fitting the 
older data quite well. The pole fit reproduces both the drop 
in $A_y$, seen in the lower energy data of Ref.~\cite{ball_P}, 
and the rapid rise displayed in the COSY data. Also shown is 
a revised fit without a pole, which takes an averaging path 
through the data without any rapid variation. 
%%%%%%%%%%%%%%%%%%%%%%%%%%%%%%%%%%%%%%%%%%%%%%%%%%%%%%
%%%%%%%%%%%%%%%%%%%%%%%%%%%%%%%%%%%%%%%%%%%%%%%%%%%%%%

In Fig.~\ref{fig:g4}, angular distributions for $A_y$ (top row)
are compared just above and below the assumed resonance energy. 
A comparison between the pole and non-pole fits at 2.38~GeV is 
less dramatic than the comparison made in Ref.~\cite{wasa_prl} 
using the SP07 prediction. Here the better data description of 
the pole fit is due to its ability to accommodate the trends 
of both the older and new data.  Given the considerable 
scatter seen in the data of Ref.~\cite{ball_P}, the need for 
a pole would gain confidence if the lower-energy data errors 
could be reduced.  

%%%%%%%%%%%%%%%%%%%%%%%%%%%%%%%%%%%%%%%%%%%%%%%%%%%%%%%%%%%%
\section{Sensitivity to other observables }
\label{sec:sensitive}
 
\begin{figure*}[th]
\centerline{
\includegraphics[height=0.35\textwidth, angle=90]{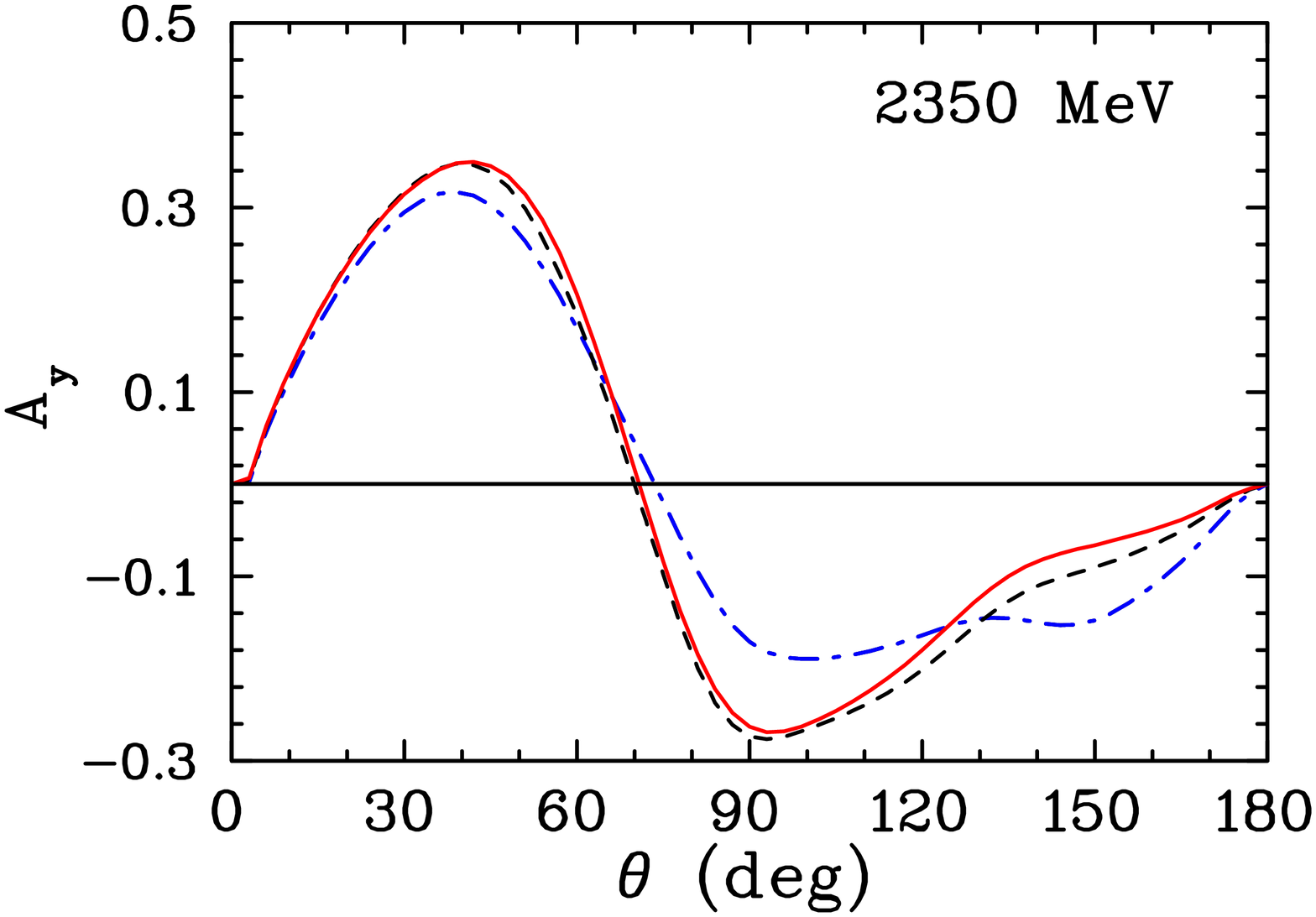}\hfill
\includegraphics[height=0.35\textwidth, angle=90]{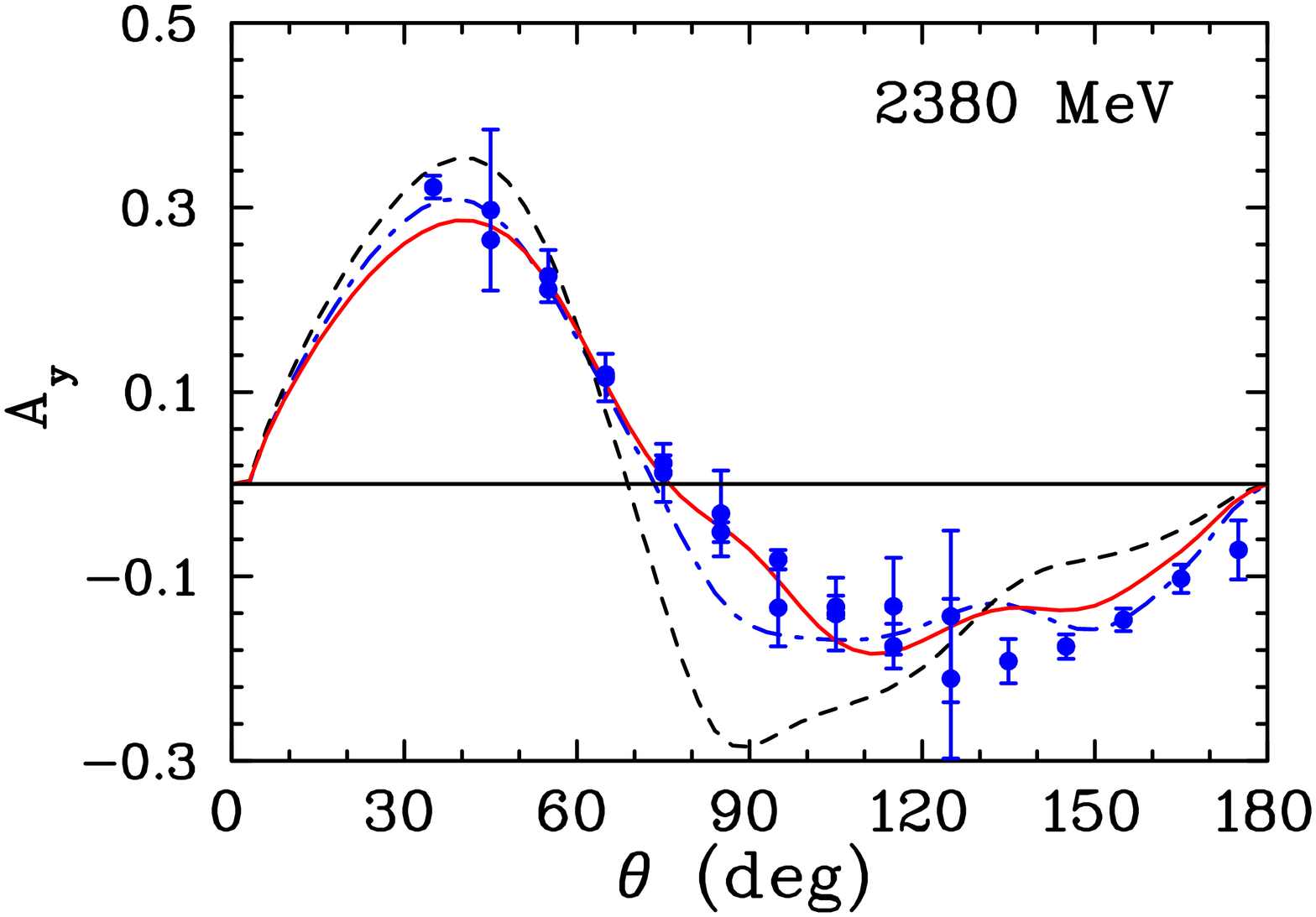}\hfill
\includegraphics[height=0.35\textwidth, angle=90]{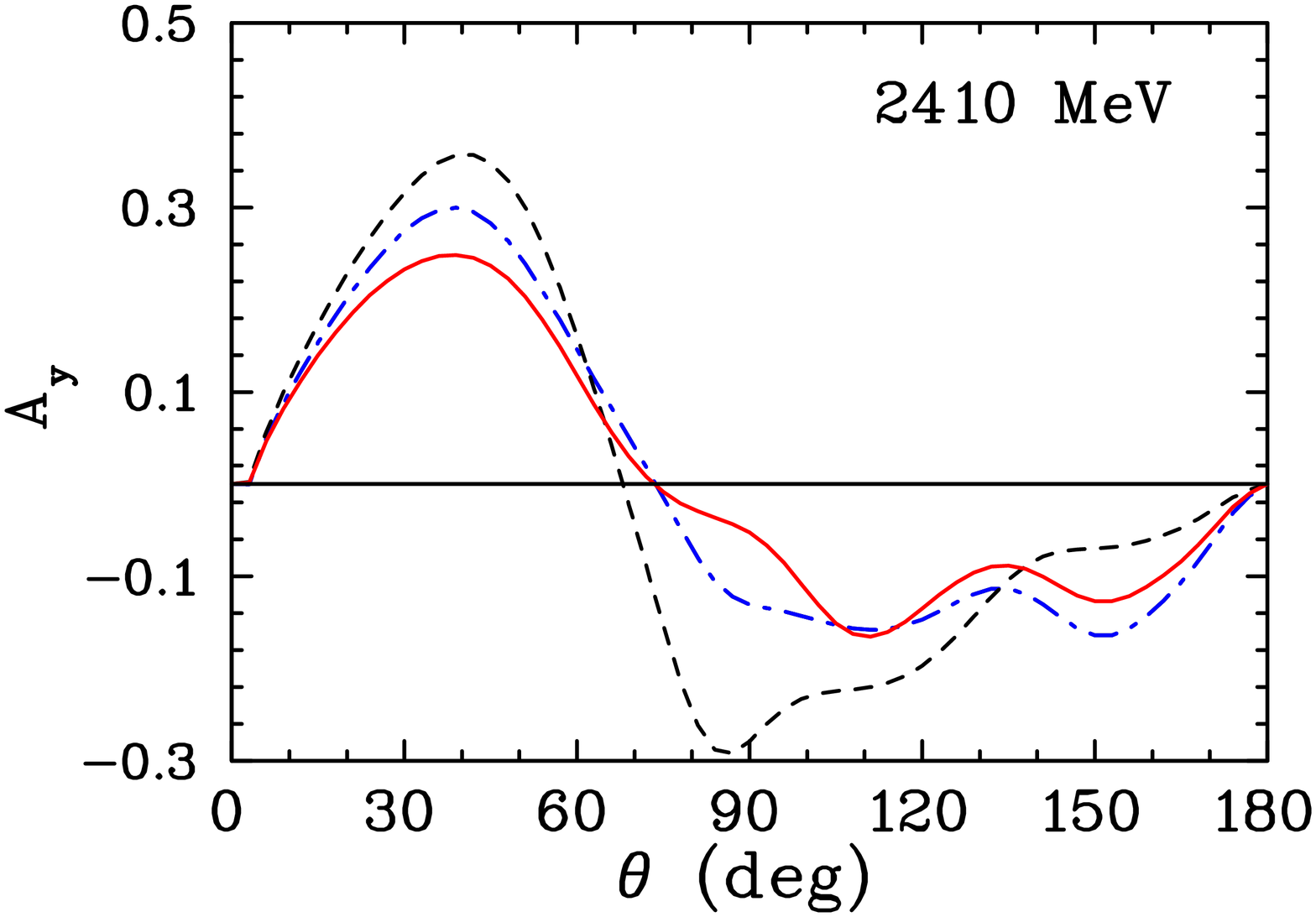}}
\centerline{
\includegraphics[height=0.35\textwidth, angle=90]{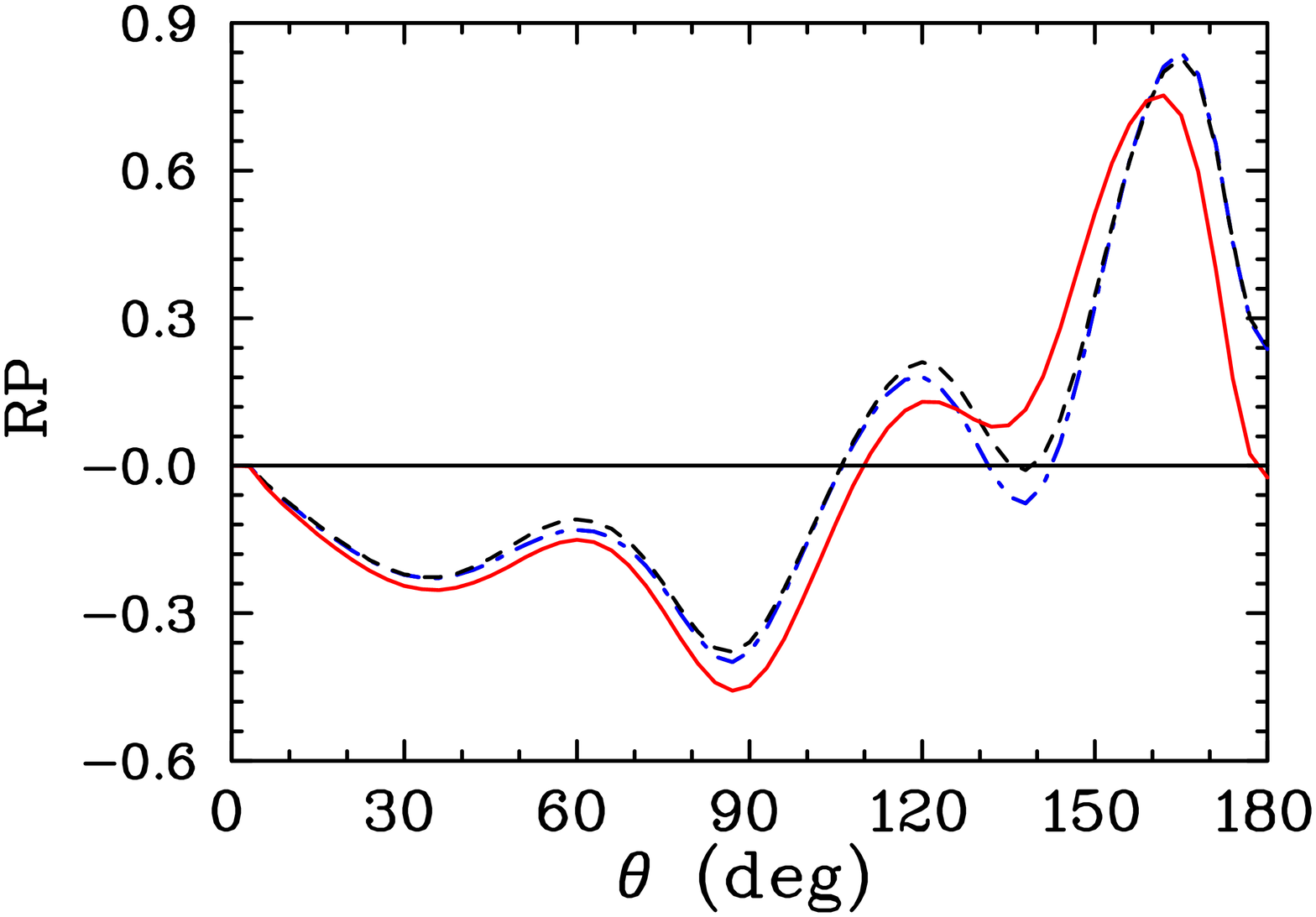}\hfill
\includegraphics[height=0.35\textwidth, angle=90]{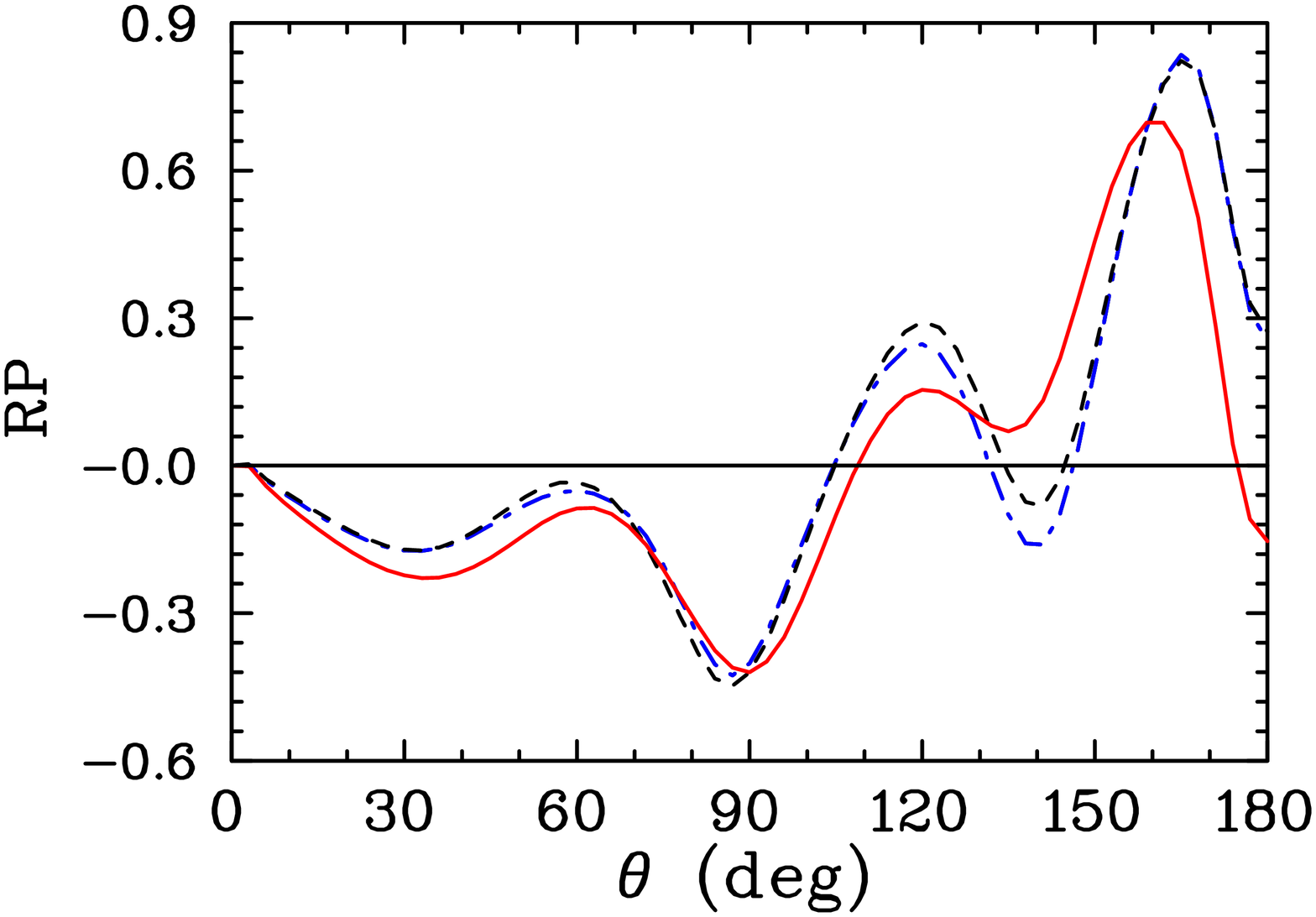}\hfill
\includegraphics[height=0.35\textwidth, angle=90]{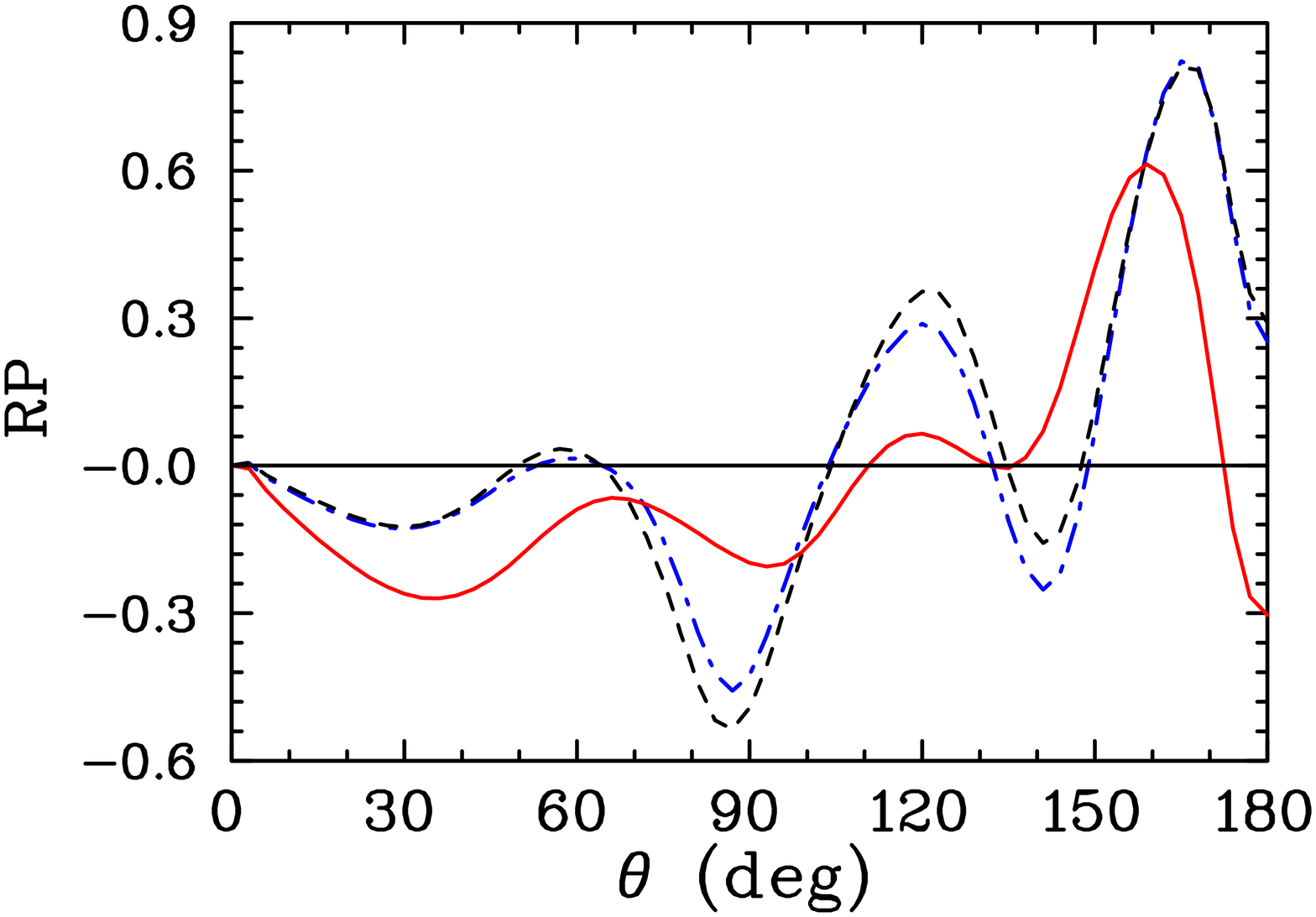}}
\centerline{
\includegraphics[height=0.35\textwidth, angle=90]{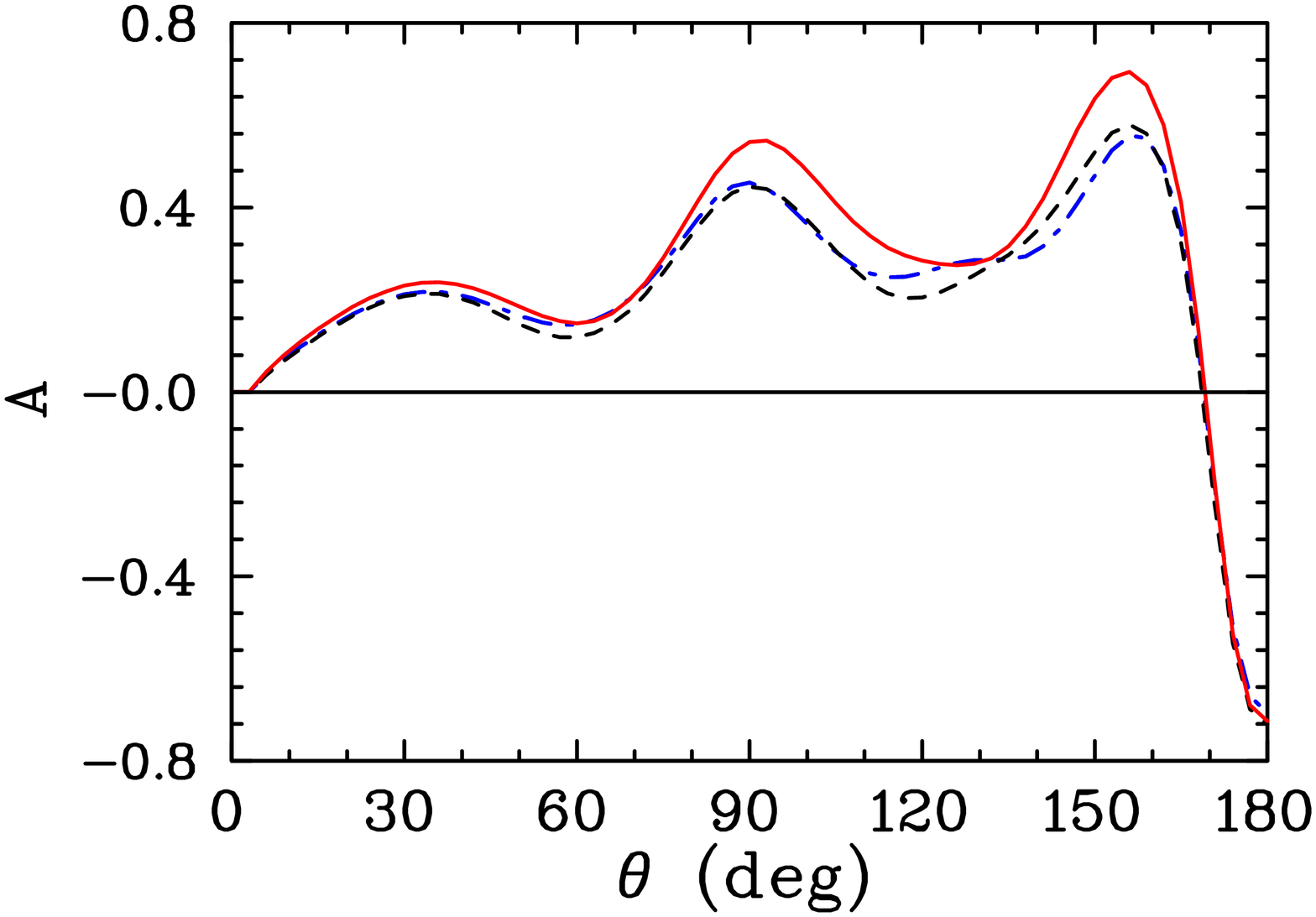}\hfill
\includegraphics[height=0.35\textwidth, angle=90]{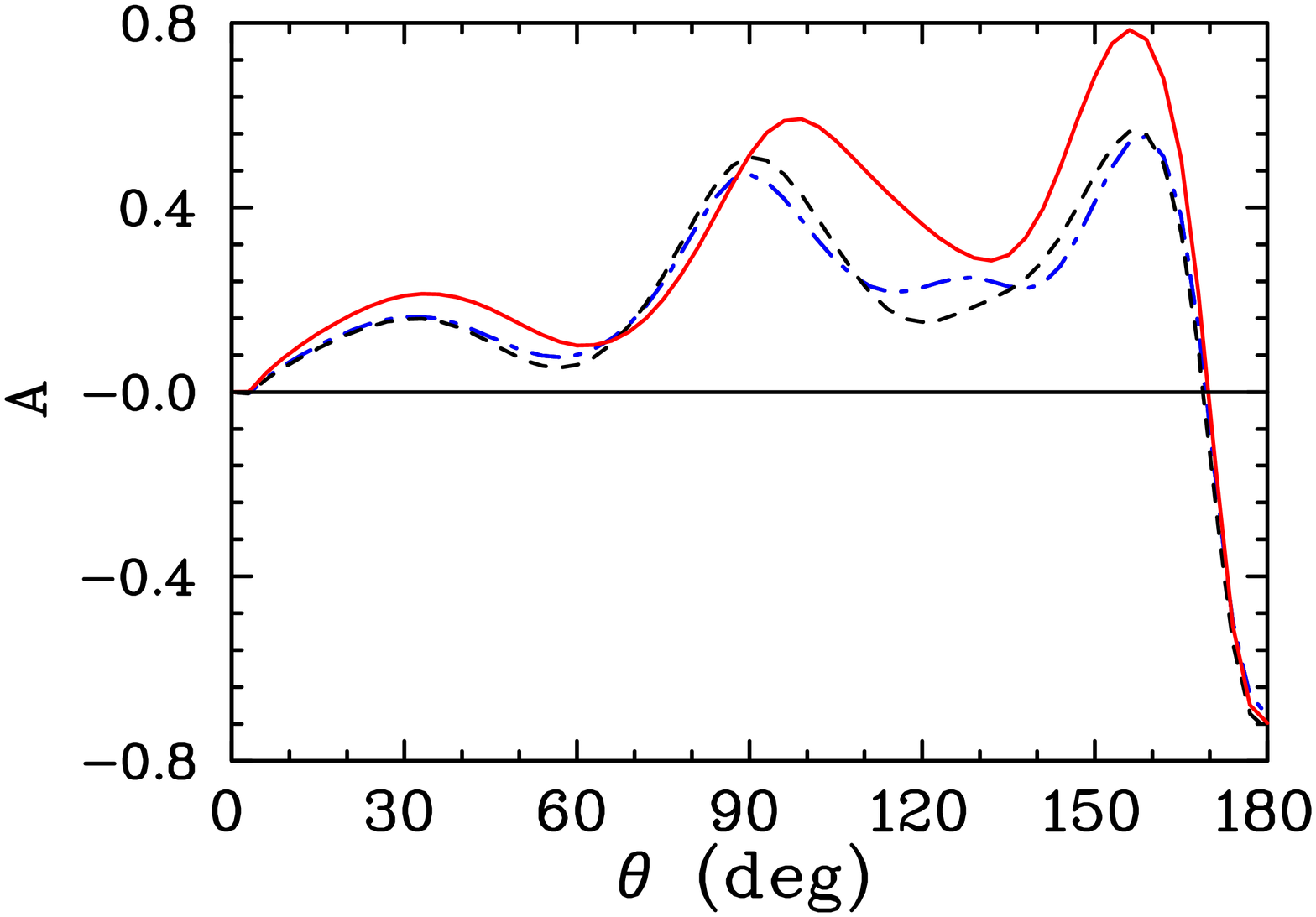}\hfill
\includegraphics[height=0.35\textwidth, angle=90]{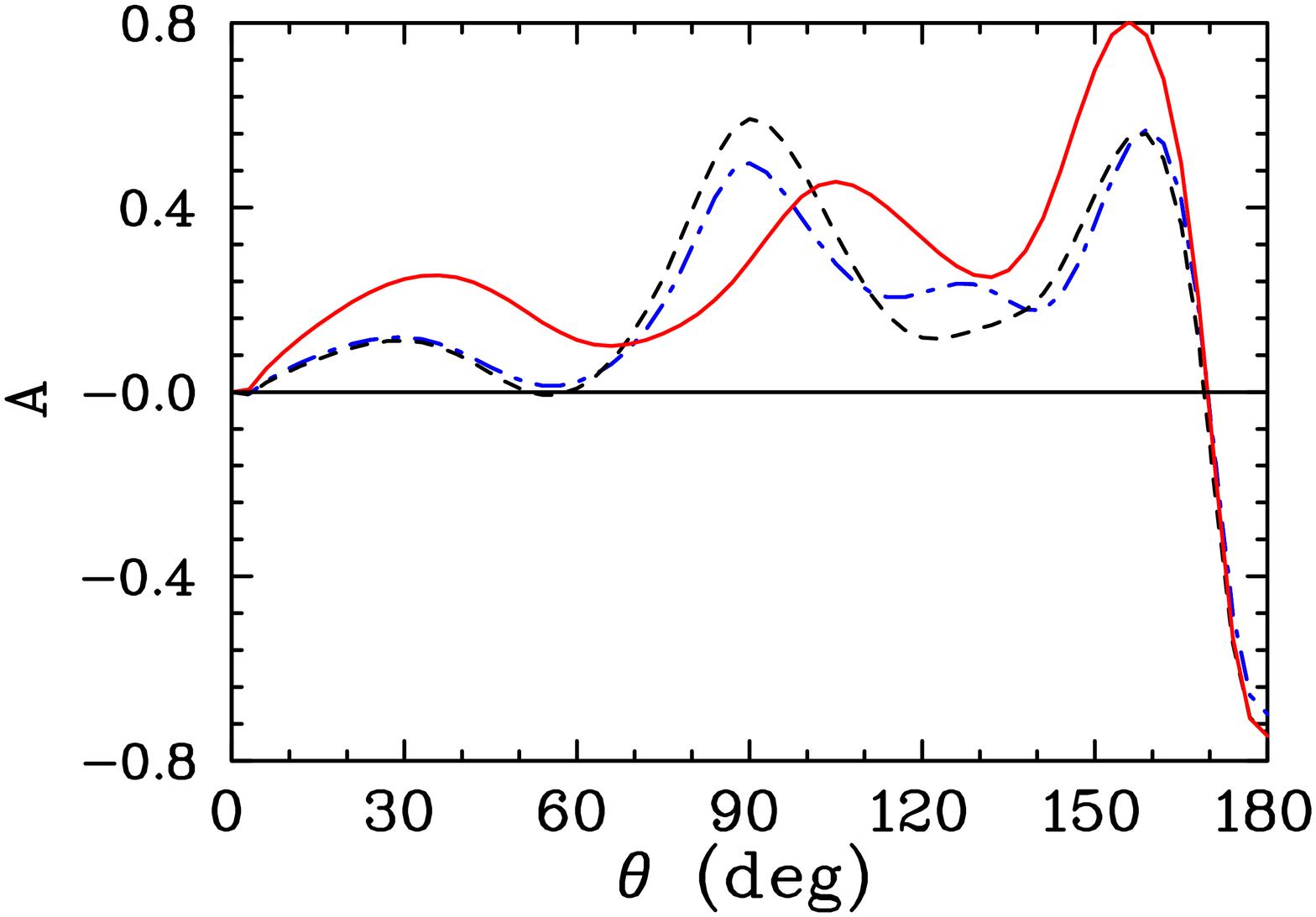}}
\vspace{3mm}
\caption{(Color online) Angular distributions of polarized observables 
	around the COSY resonance~\protect\cite{wasa_prl}: W = 2350~MeV
        (left panels), 2380~MeV (middle panels), and 2410~MeV 
	(right panels). Previous measurements within $\Delta W 
	=\pm 5~MeV$~\protect\cite{SAID} are plotted. Notation as in 
	Fig.~\protect\ref{fig:g3}. \label{fig:g4}}
\end{figure*}

\begin{figure*}[th]
\centerline{
\includegraphics[height=0.35\textwidth, angle=90]{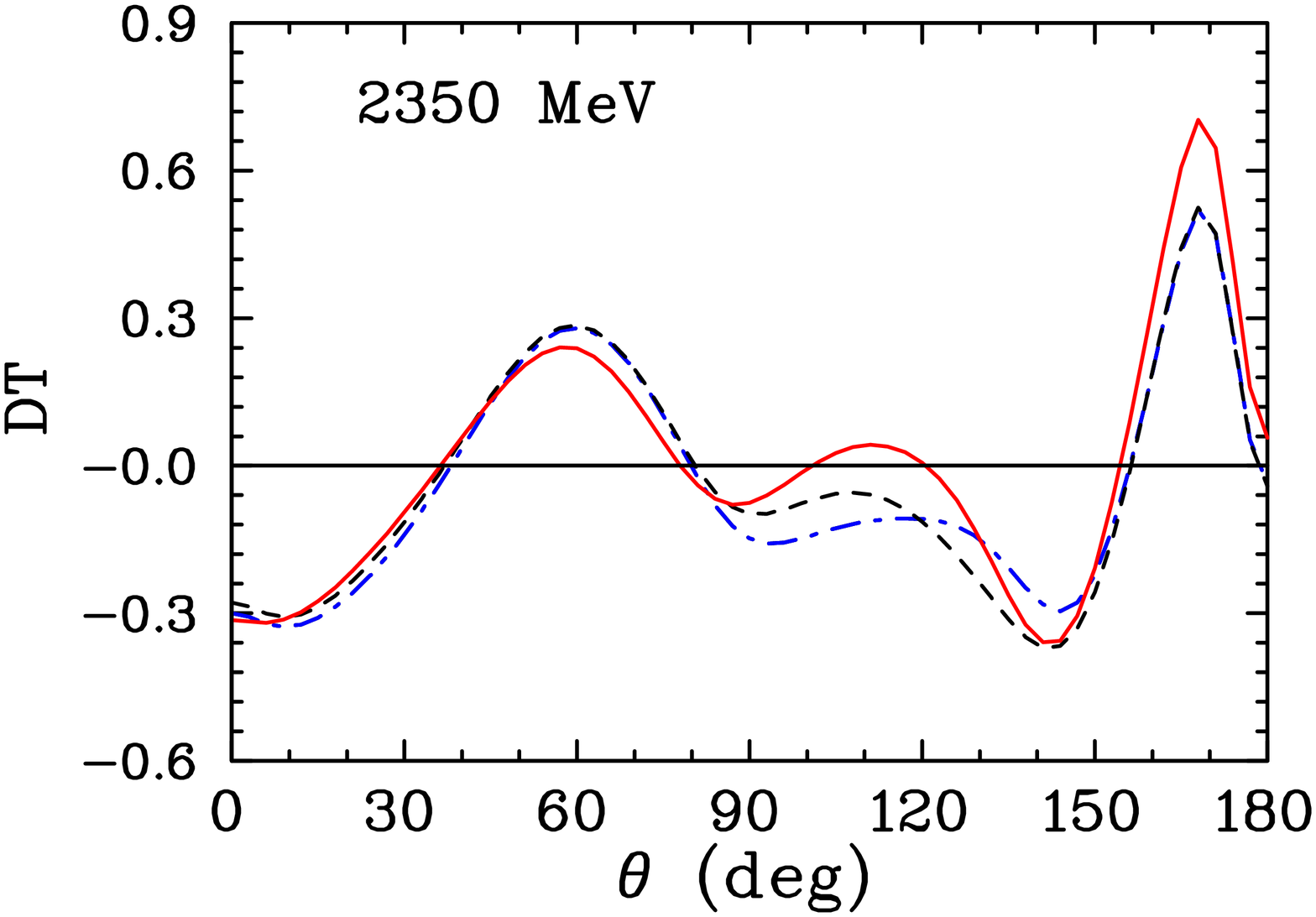}\hfill
\includegraphics[height=0.35\textwidth, angle=90]{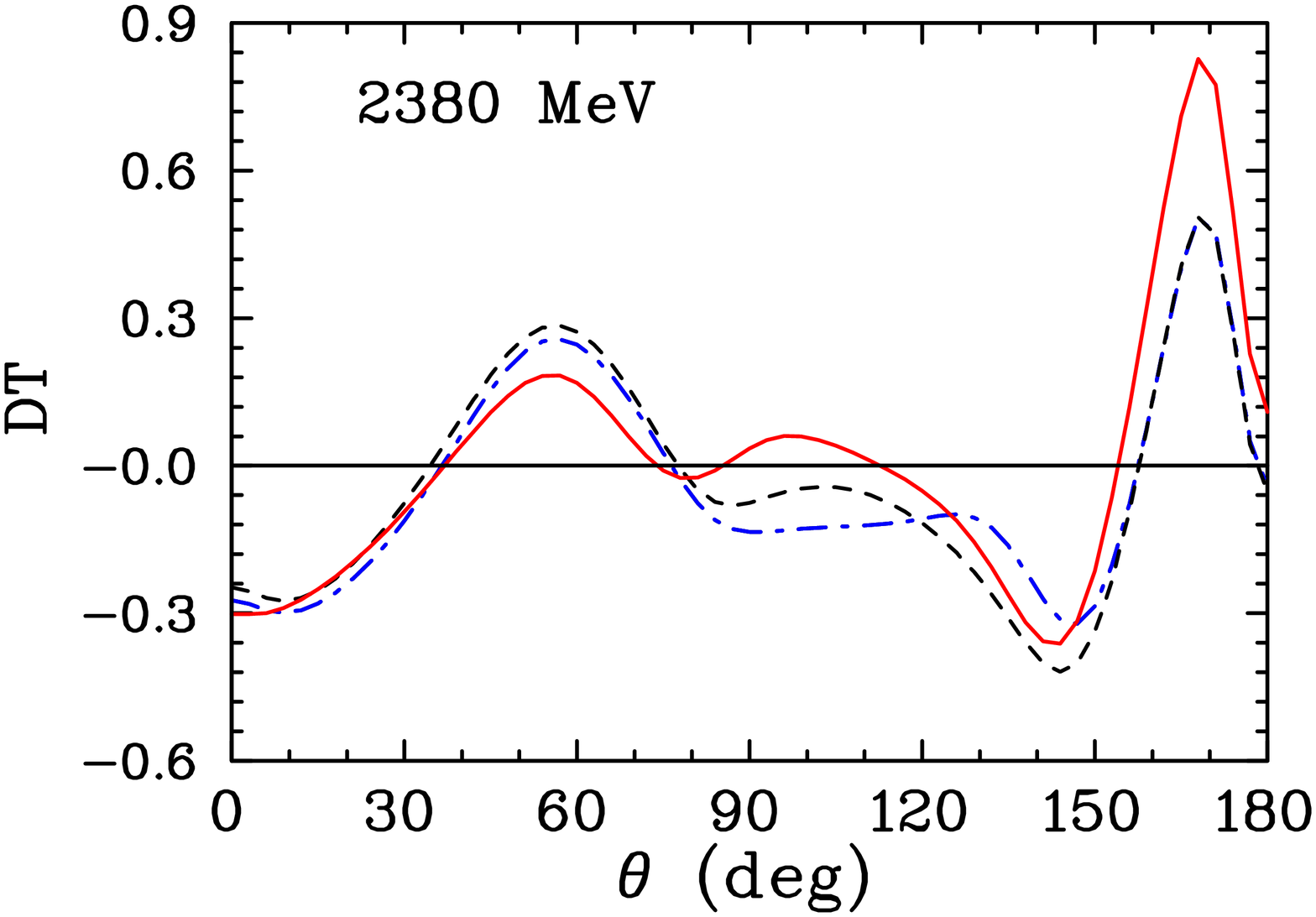}\hfill
\includegraphics[height=0.35\textwidth, angle=90]{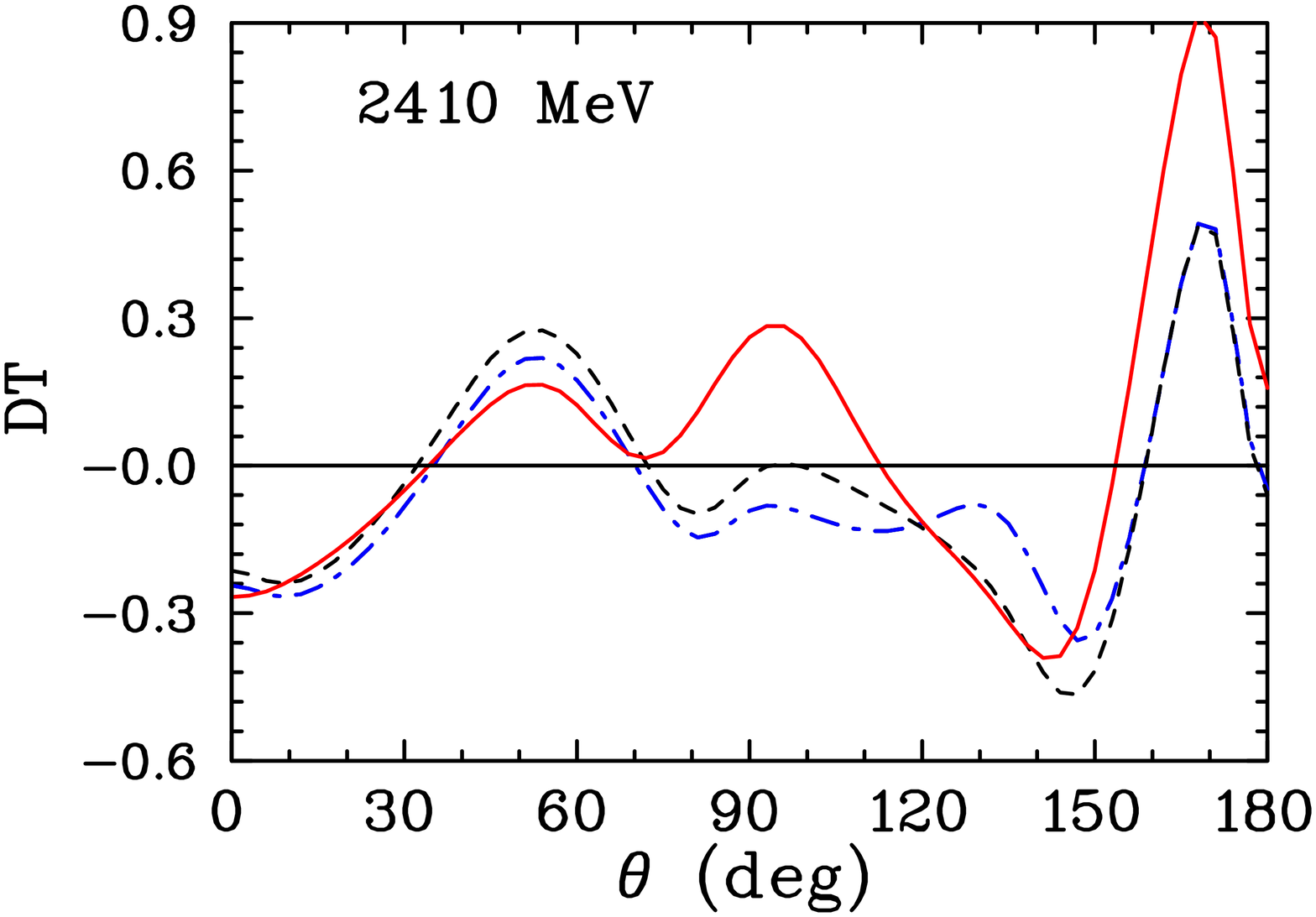}}
\centerline{
\includegraphics[height=0.35\textwidth, angle=90]{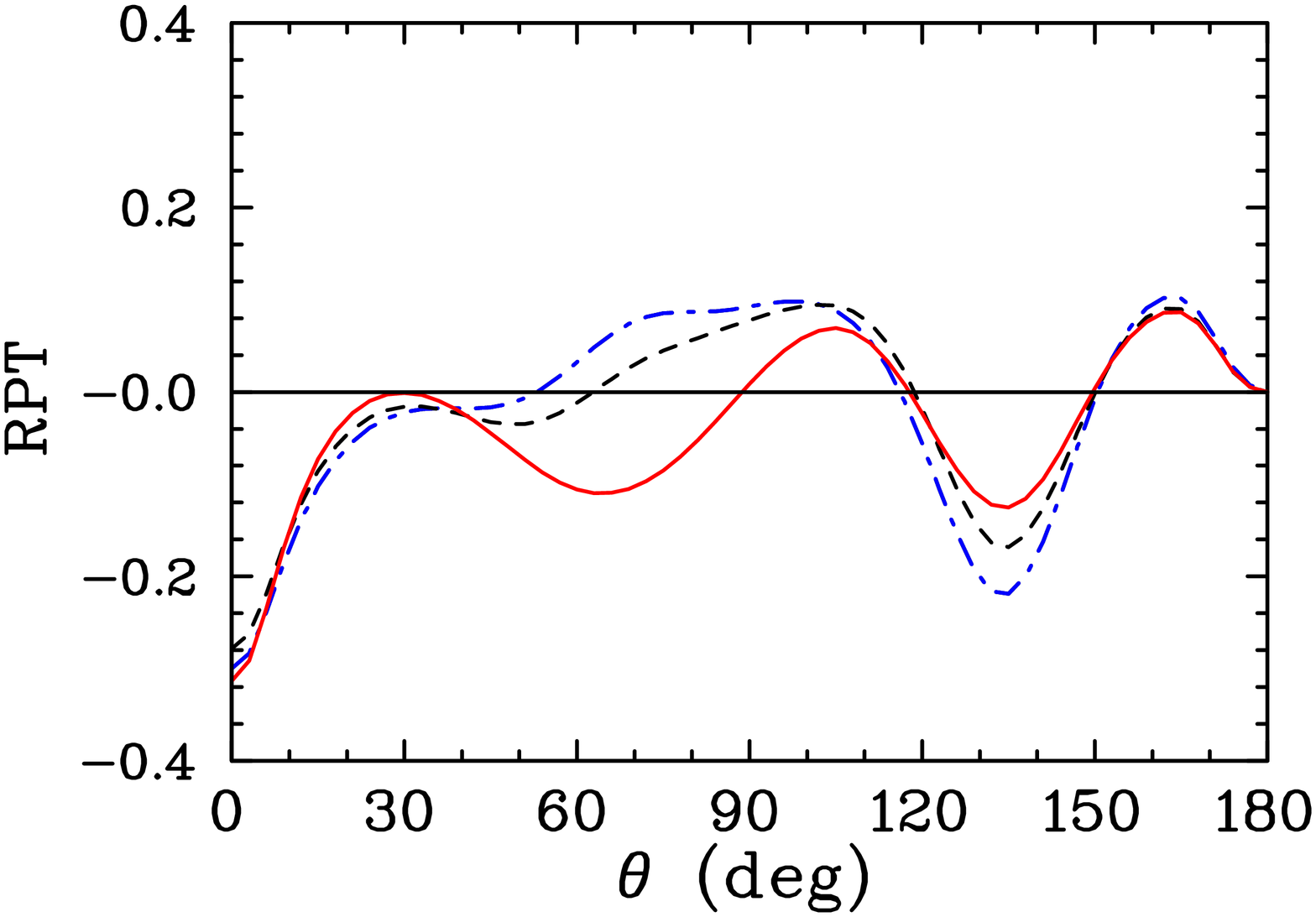}\hfill
\includegraphics[height=0.35\textwidth, angle=90]{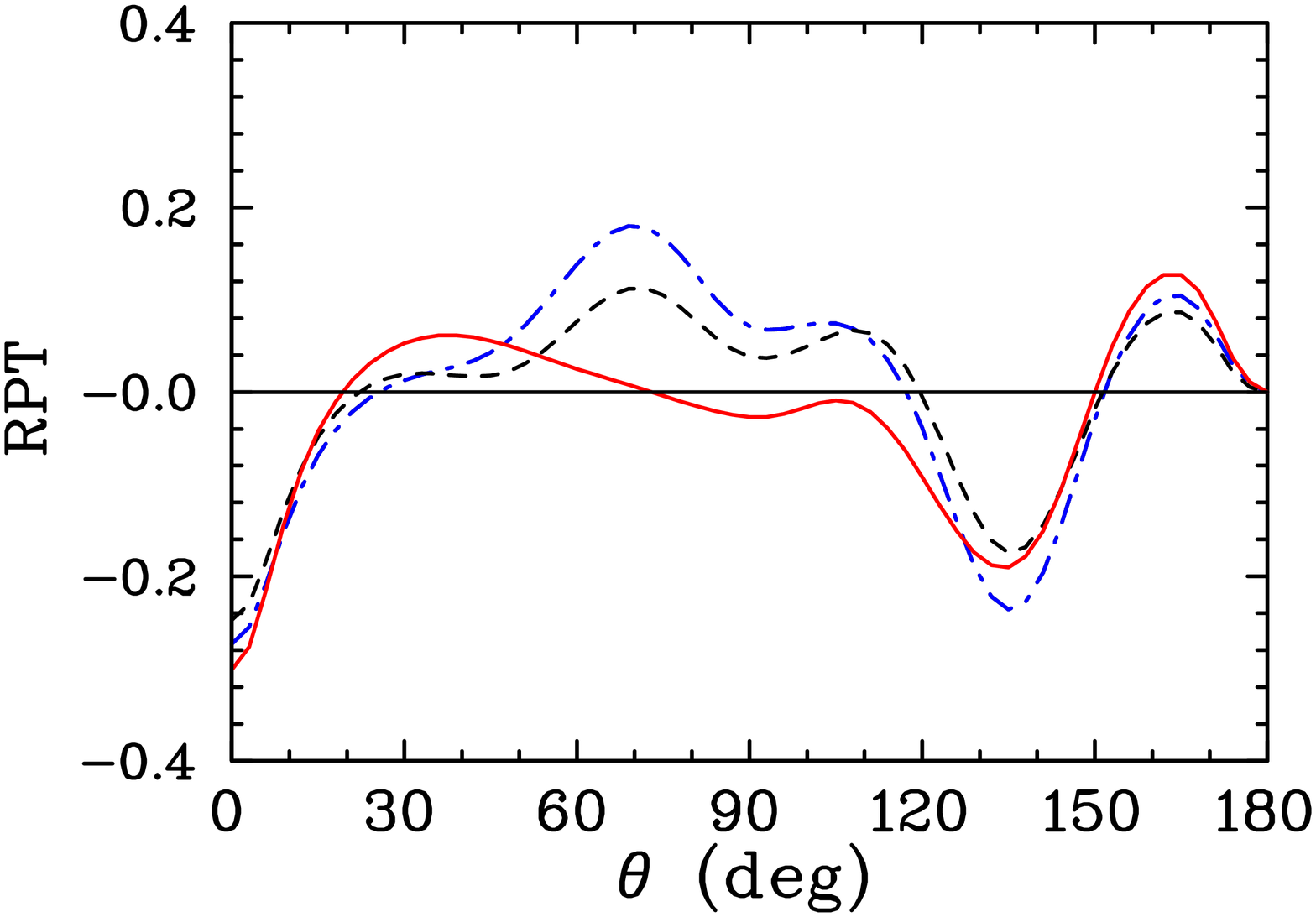}\hfill
\includegraphics[height=0.35\textwidth, angle=90]{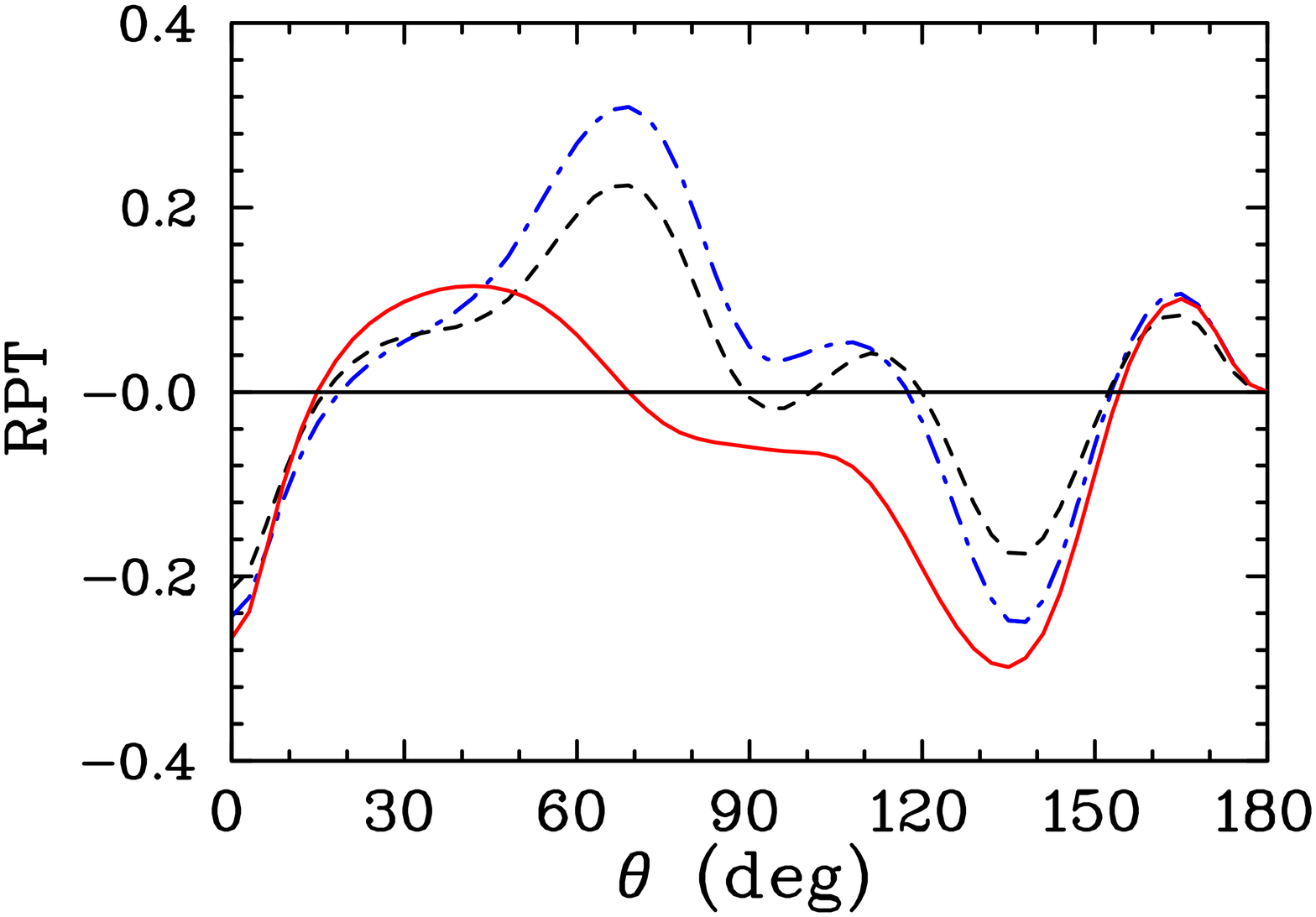}}
\centerline{
\includegraphics[height=0.35\textwidth, angle=90]{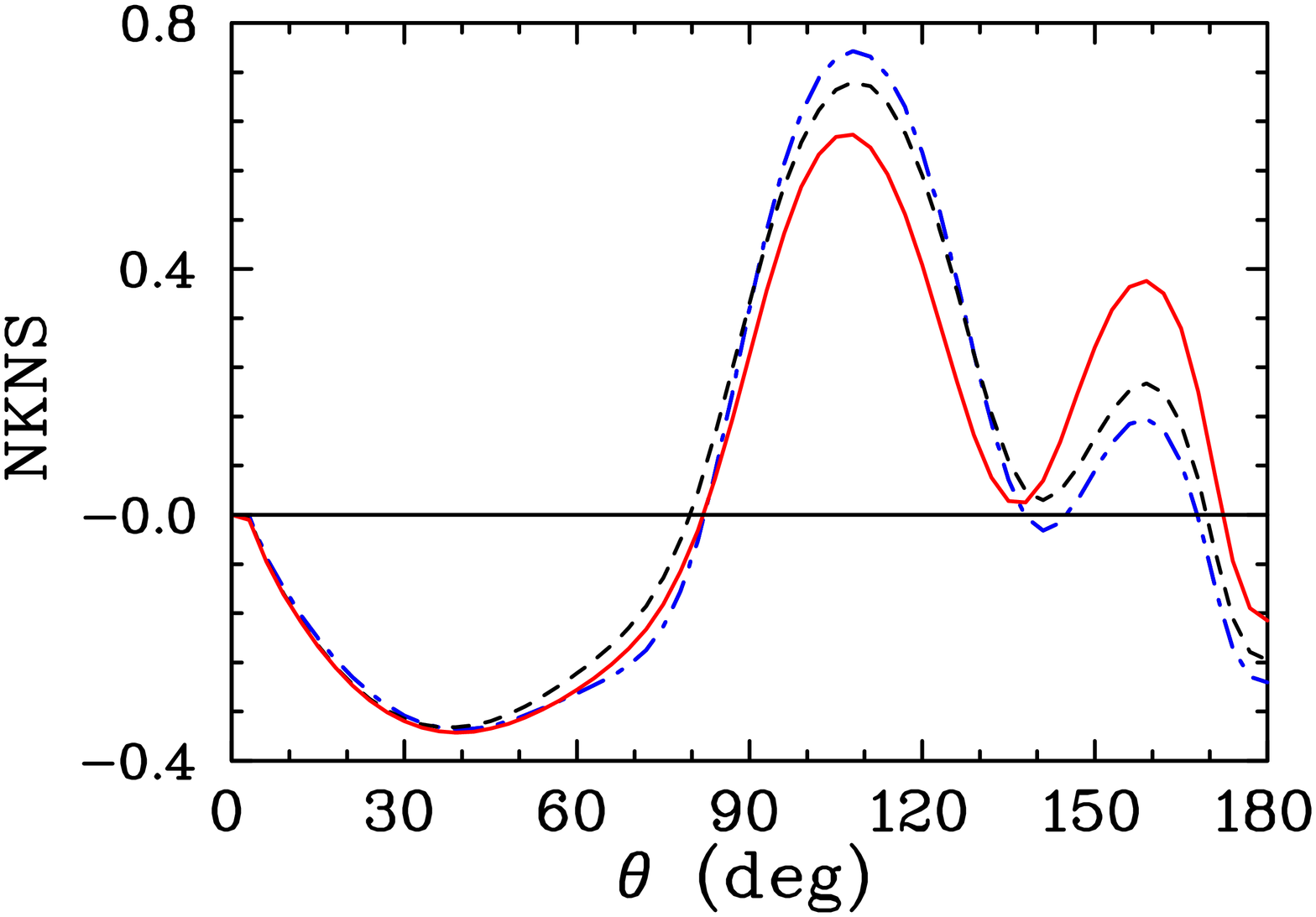}\hfill
\includegraphics[height=0.35\textwidth, angle=90]{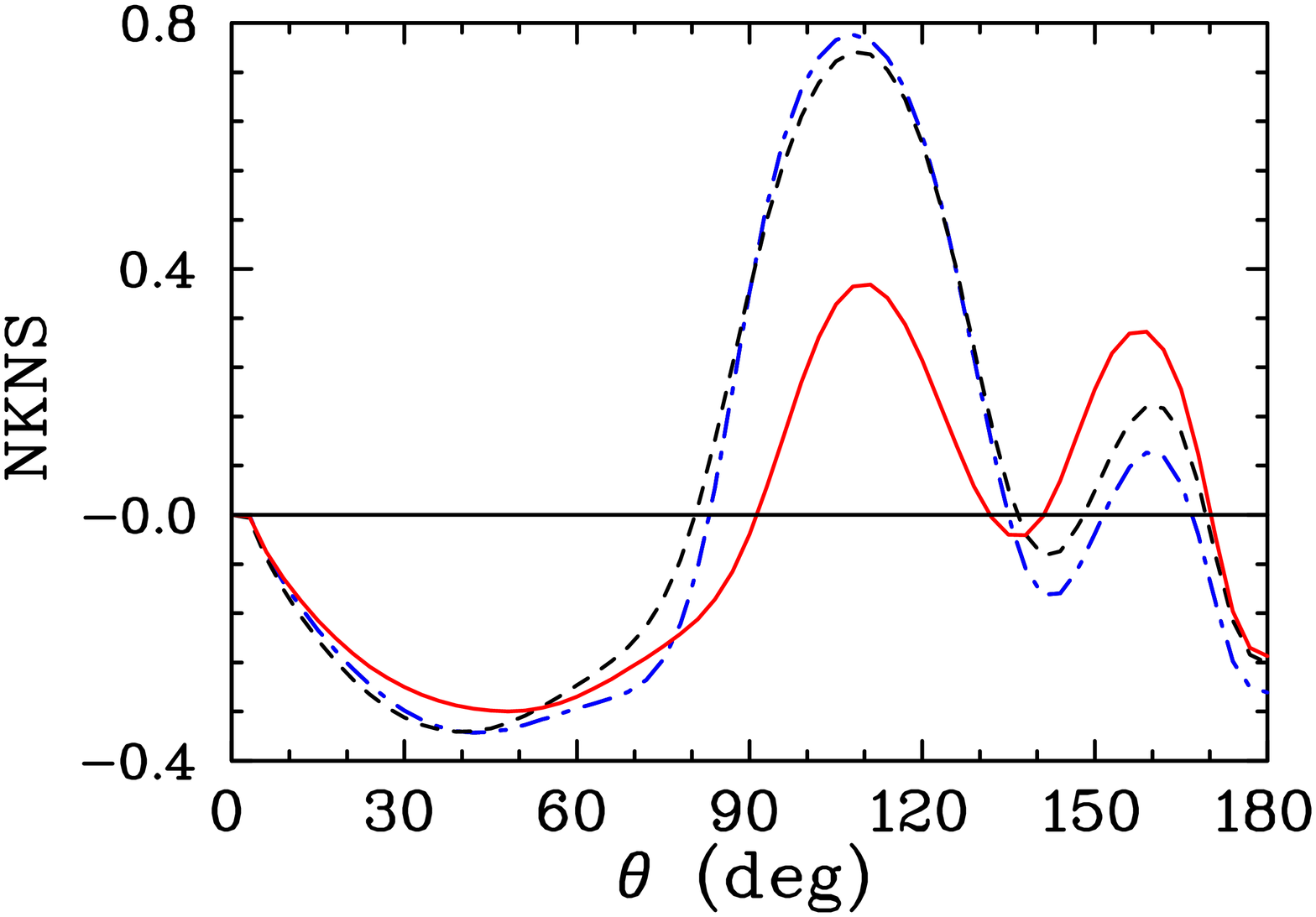}\hfill
\includegraphics[height=0.35\textwidth, angle=90]{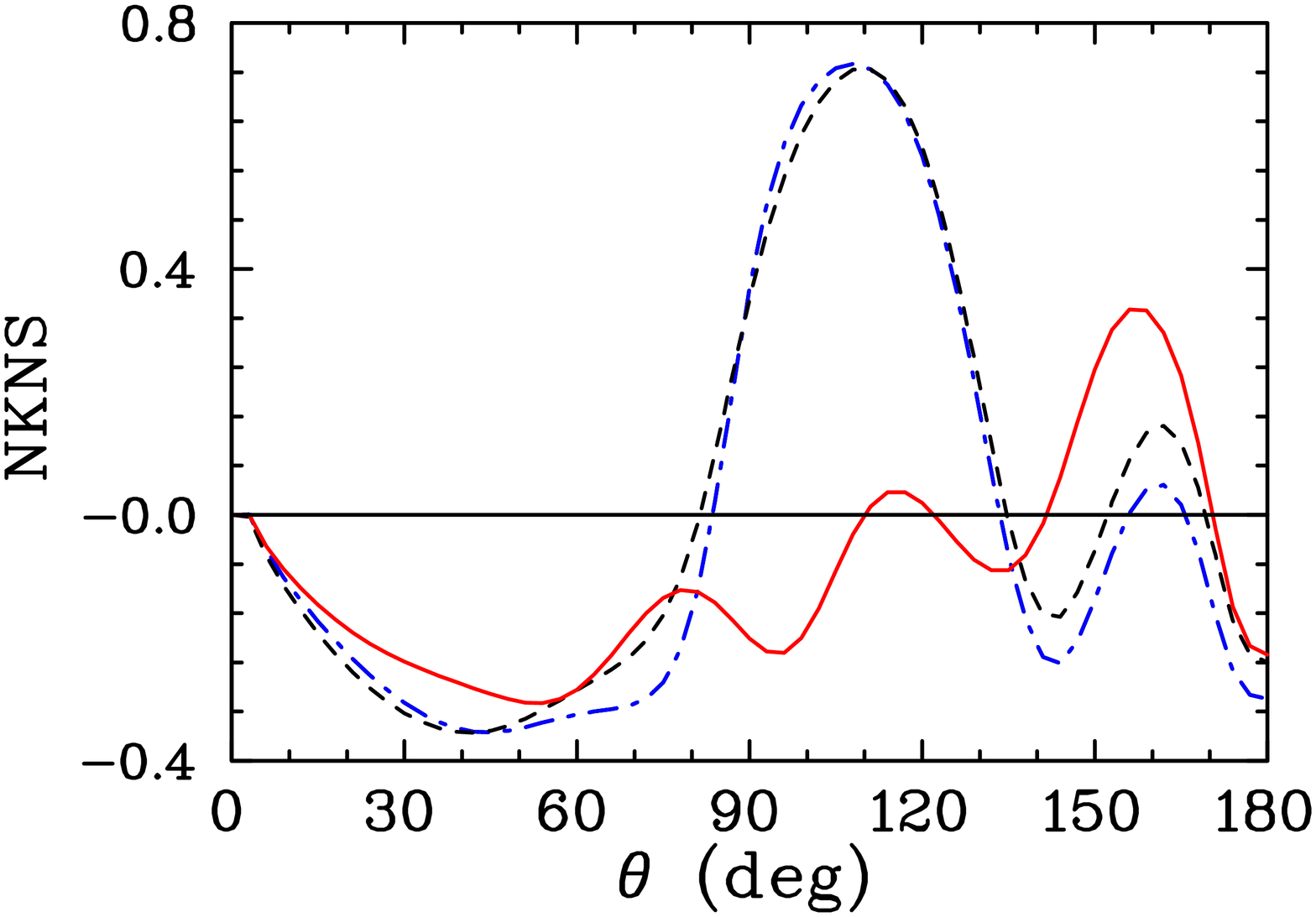}}
\vspace{3mm}
\caption{(Color online) Angular distributions of polarized observables 
	around the COSY resonance~\protect\cite{wasa_prl}: W = 2350~MeV 
	(left panels), 2380~MeV (middle panels), and 2410~MeV 
	(right panels). Notation as in Fig.~\protect\ref{fig:g3}. 
	\label{fig:g5}}
\end{figure*}

\begin{figure*}[th]
\centerline{
\includegraphics[height=0.35\textwidth, angle=90]{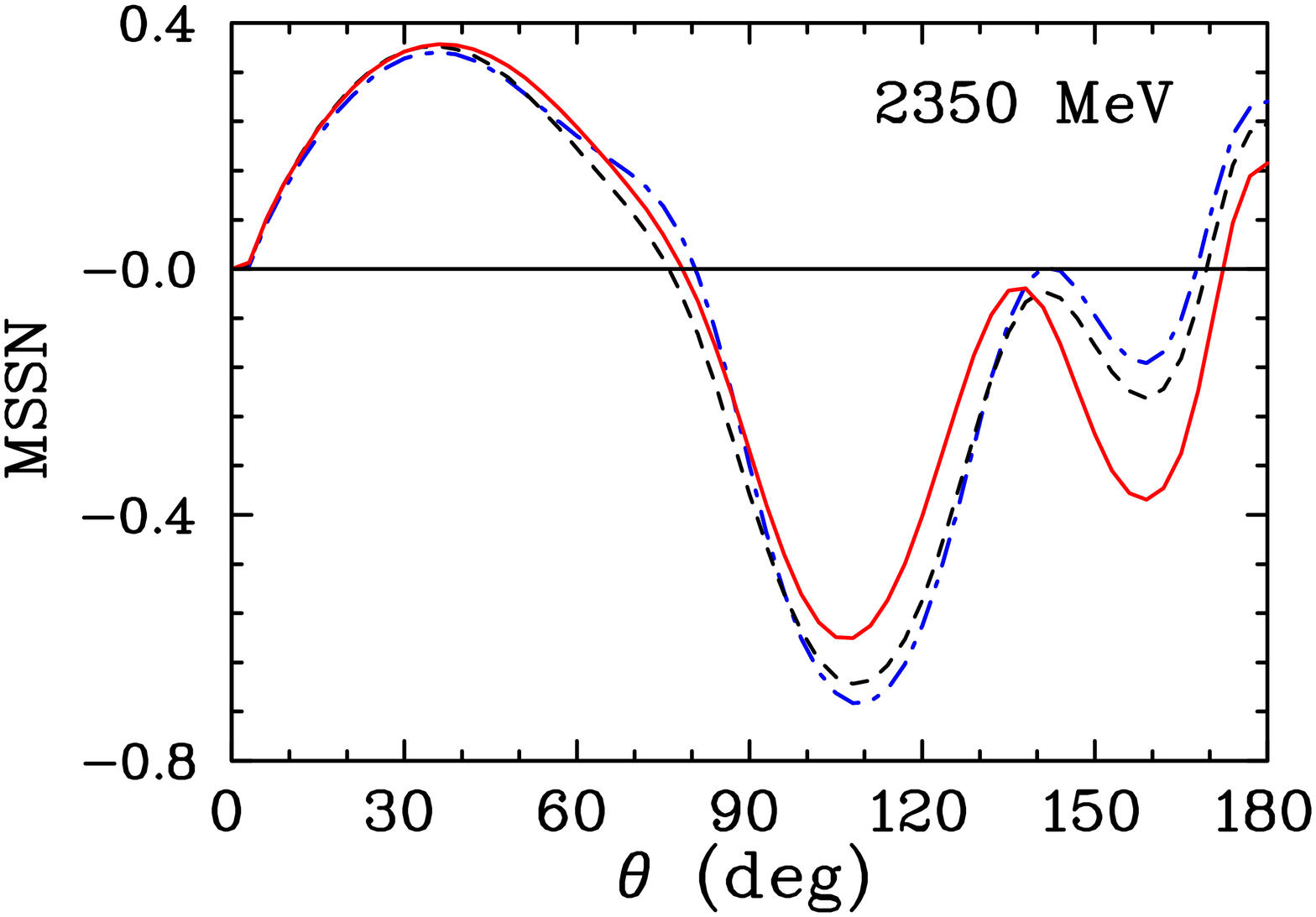}\hfill
\includegraphics[height=0.35\textwidth, angle=90]{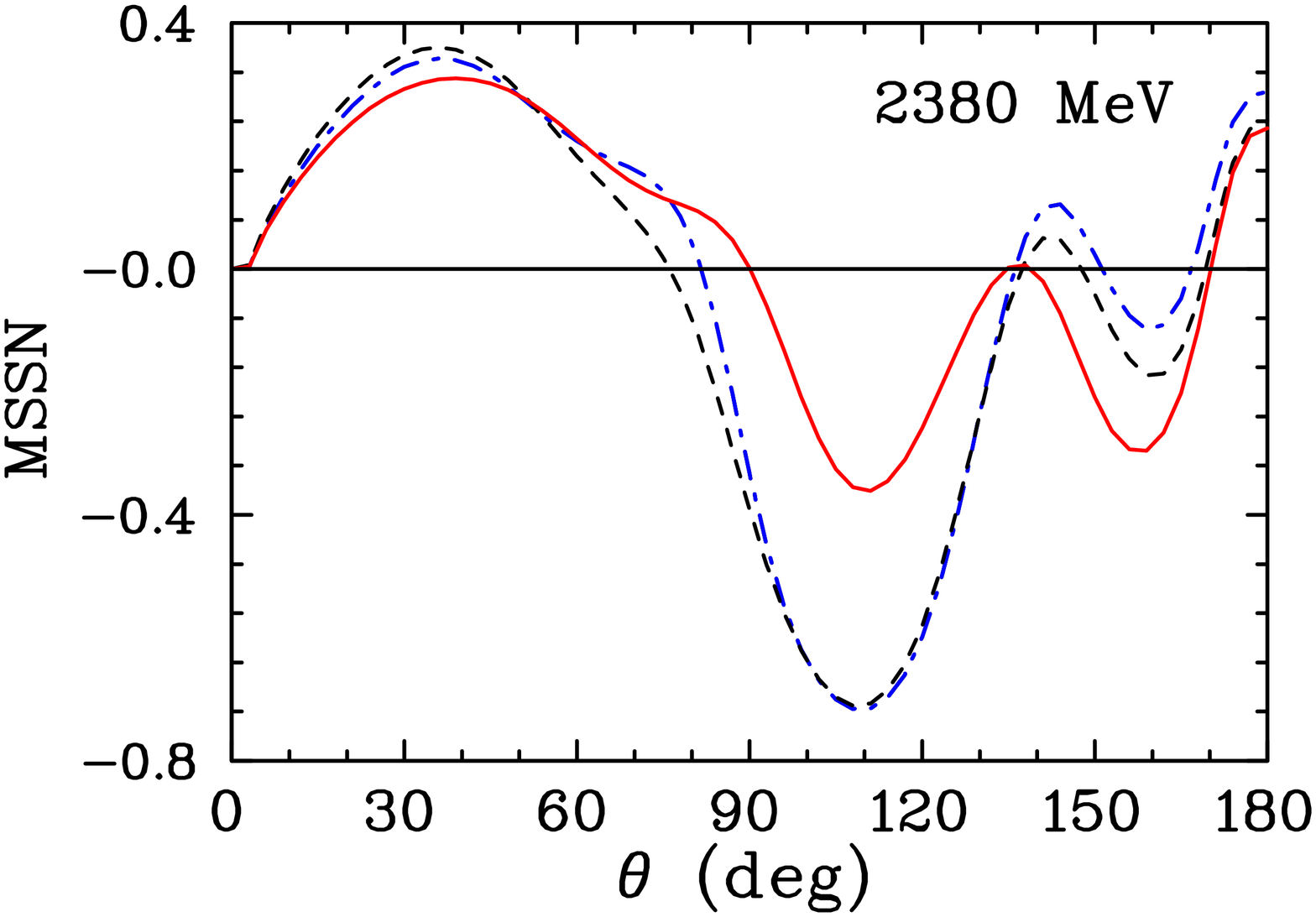}\hfill
\includegraphics[height=0.35\textwidth, angle=90]{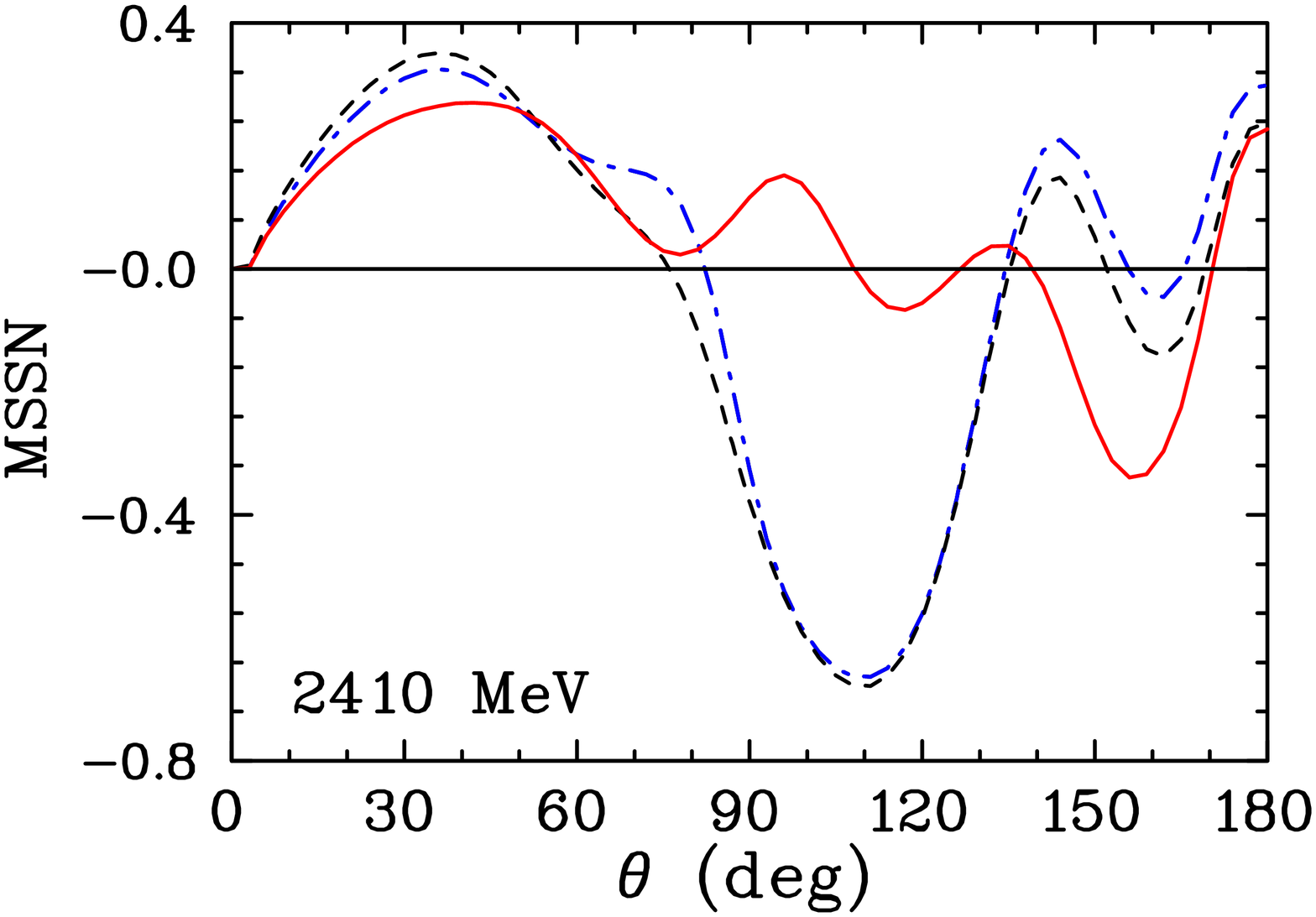}}
\centerline{
\includegraphics[height=0.35\textwidth, angle=90]{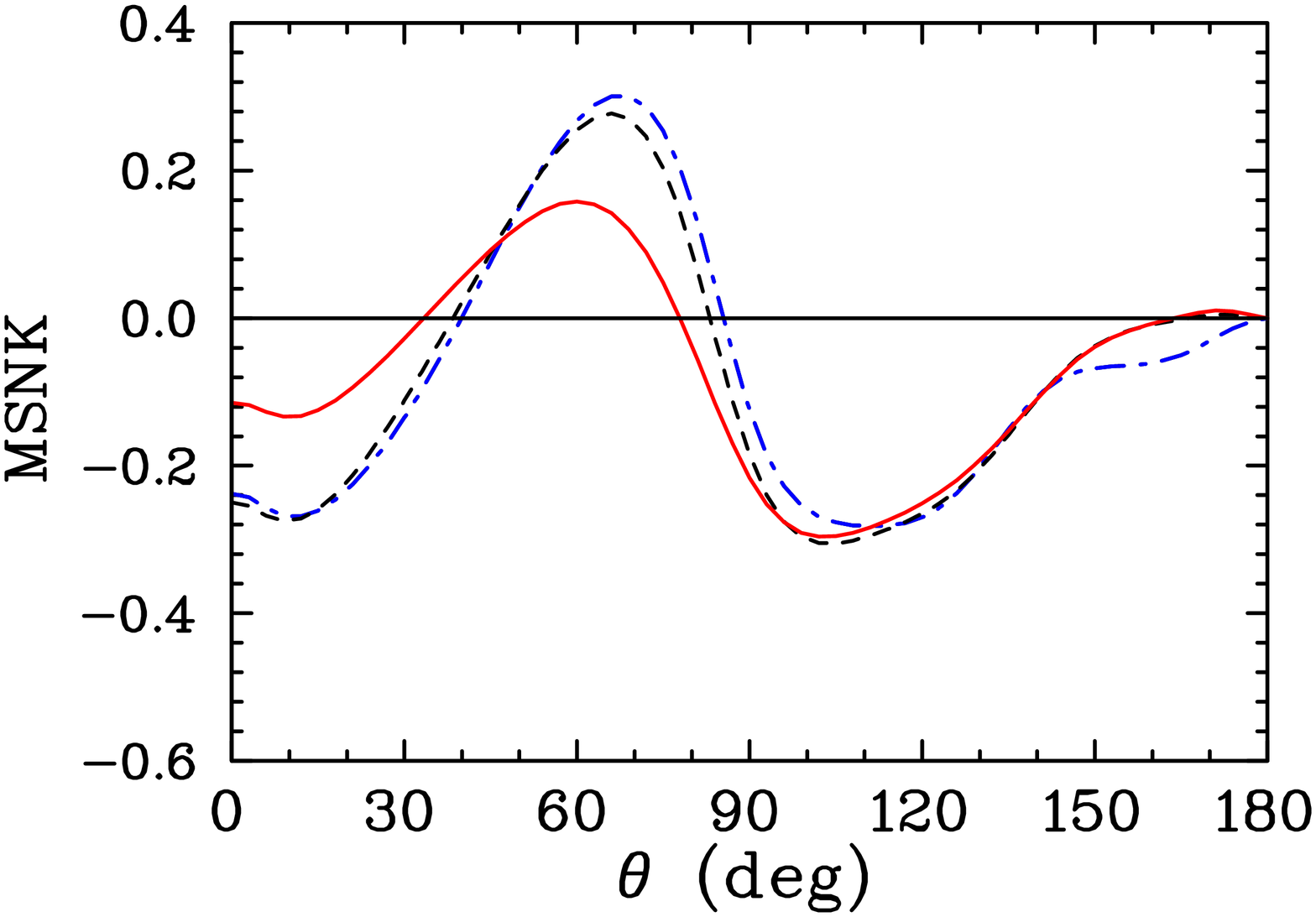}\hfill
\includegraphics[height=0.35\textwidth, angle=90]{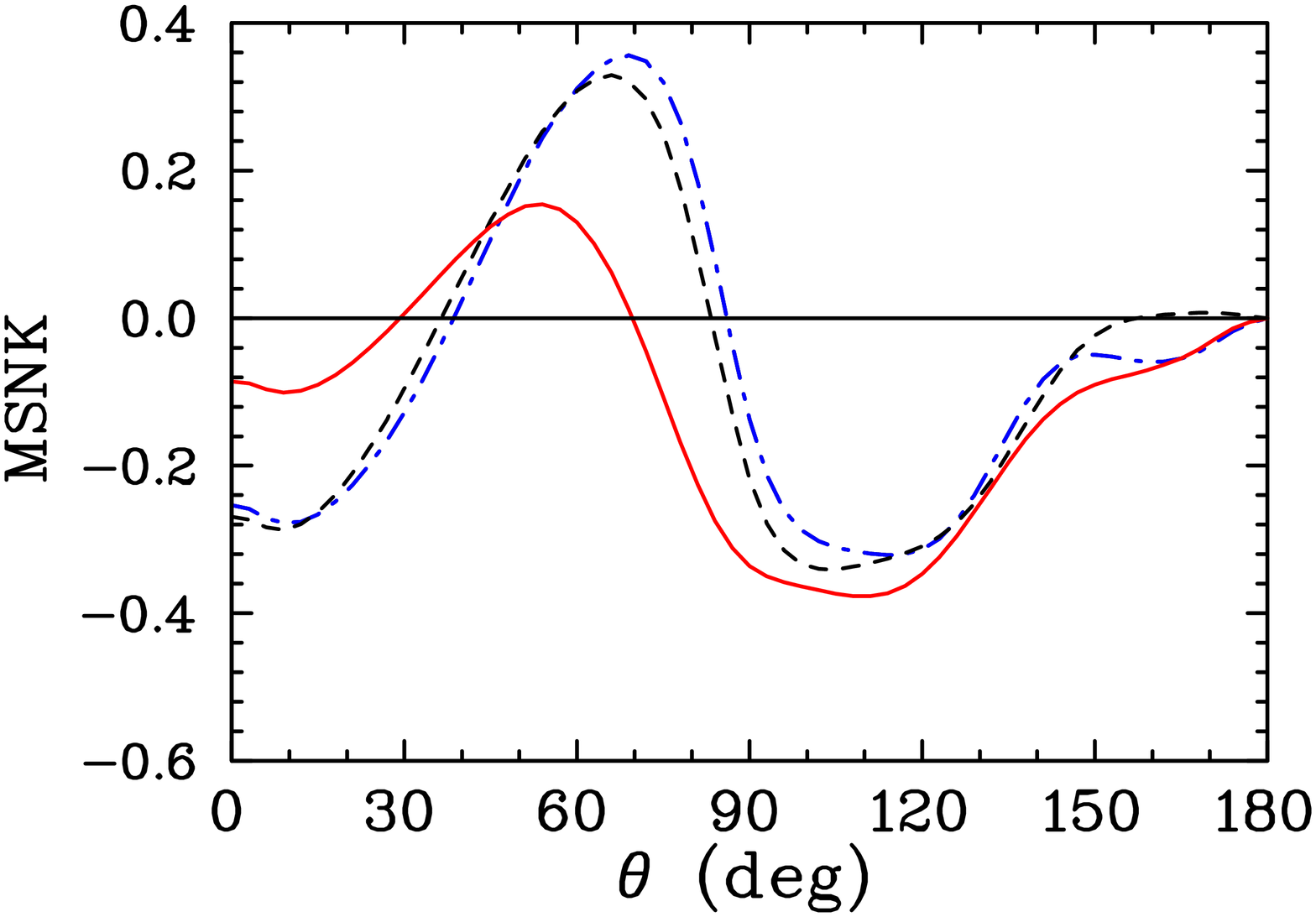}\hfill
\includegraphics[height=0.35\textwidth, angle=90]{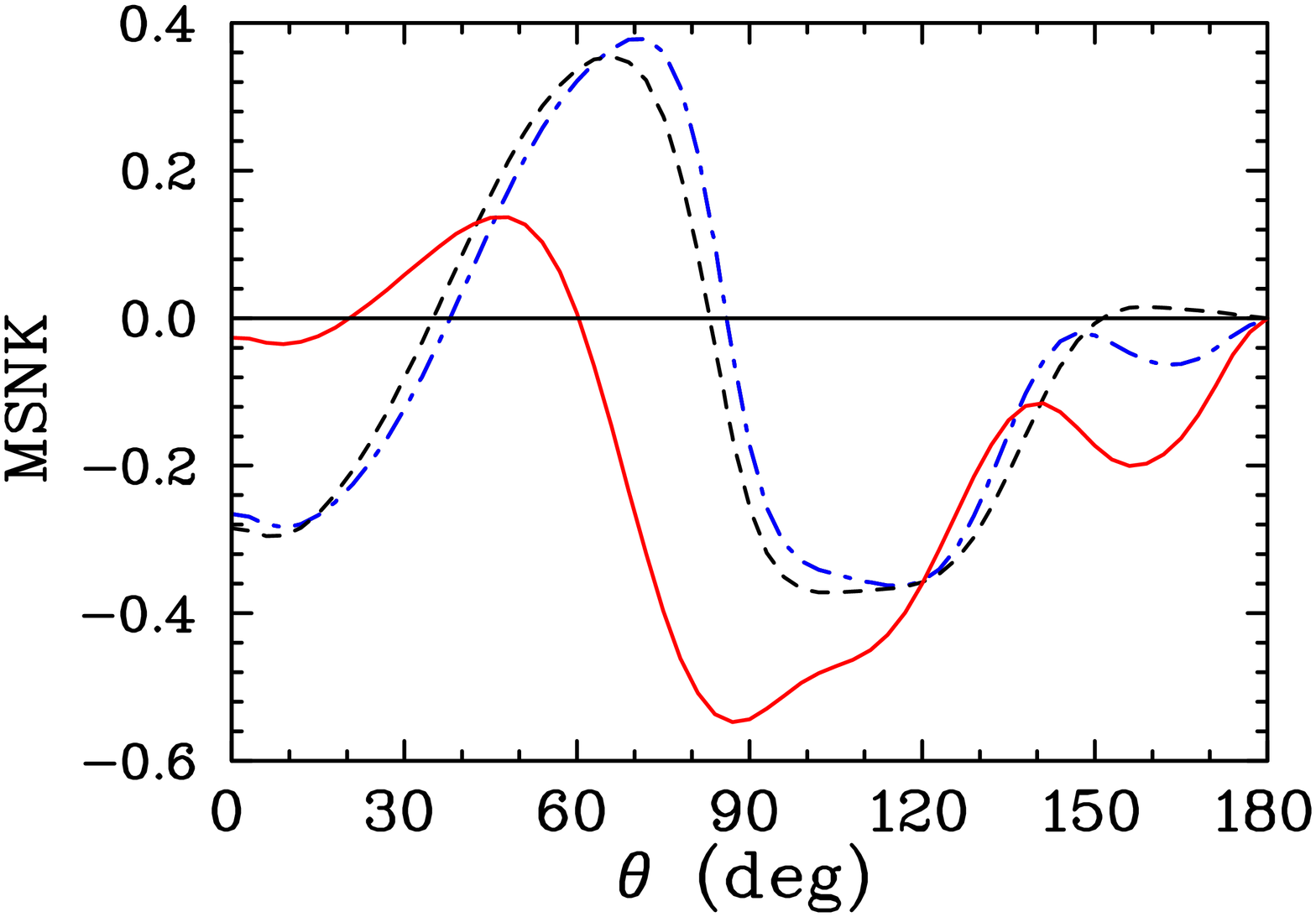}}
\centerline{
\includegraphics[height=0.35\textwidth, angle=90]{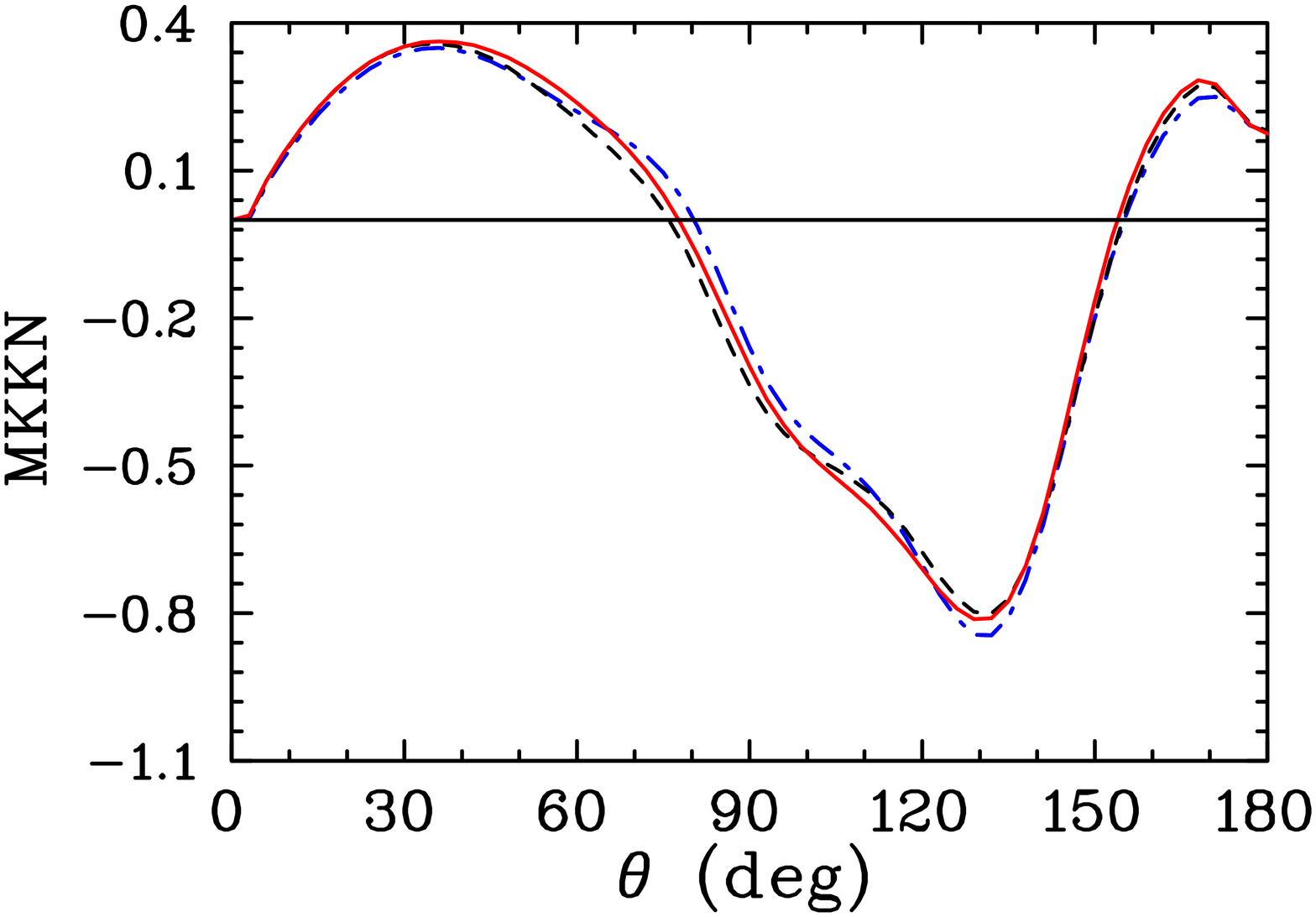}\hfill
\includegraphics[height=0.35\textwidth, angle=90]{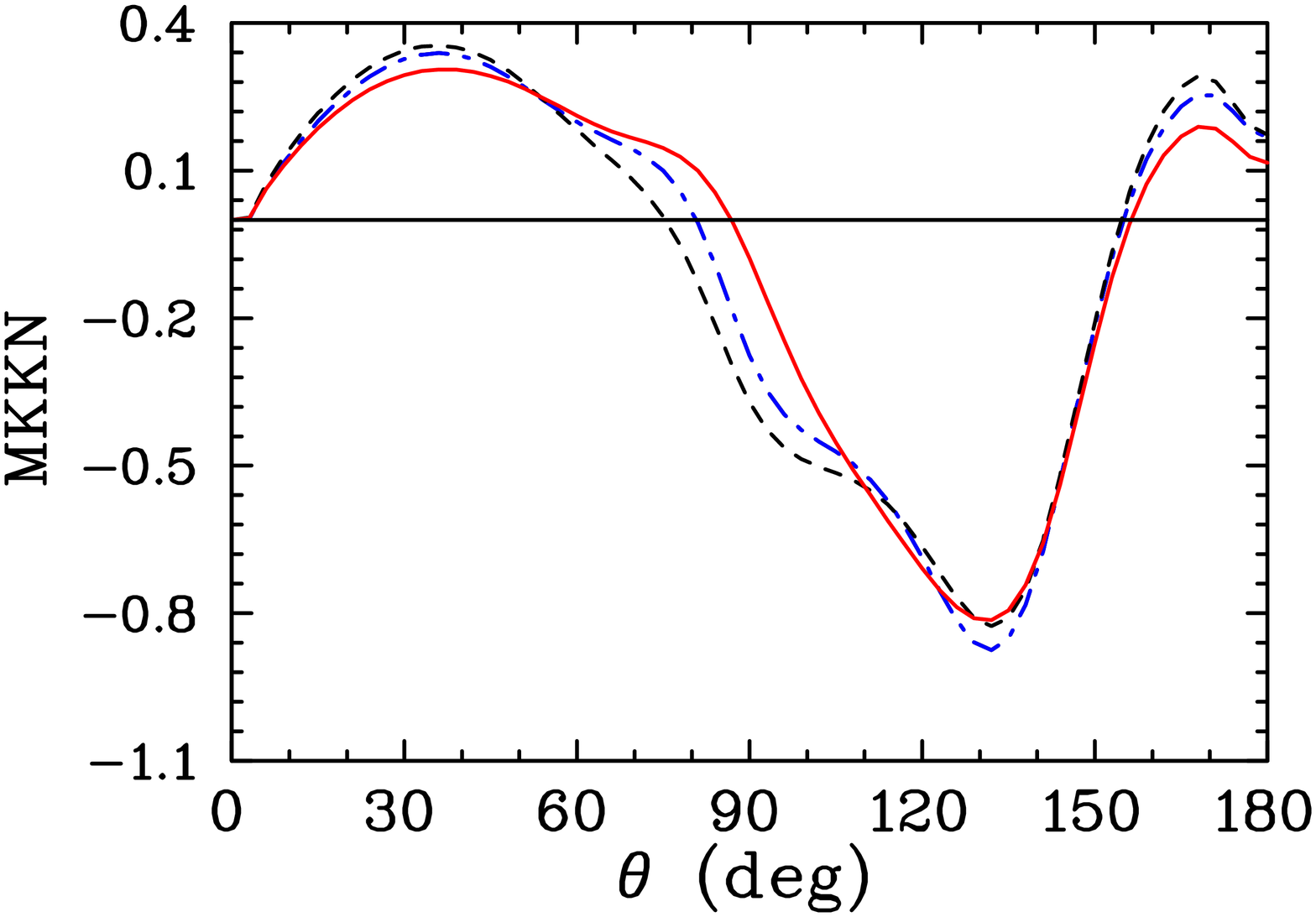}\hfill
\includegraphics[height=0.35\textwidth, angle=90]{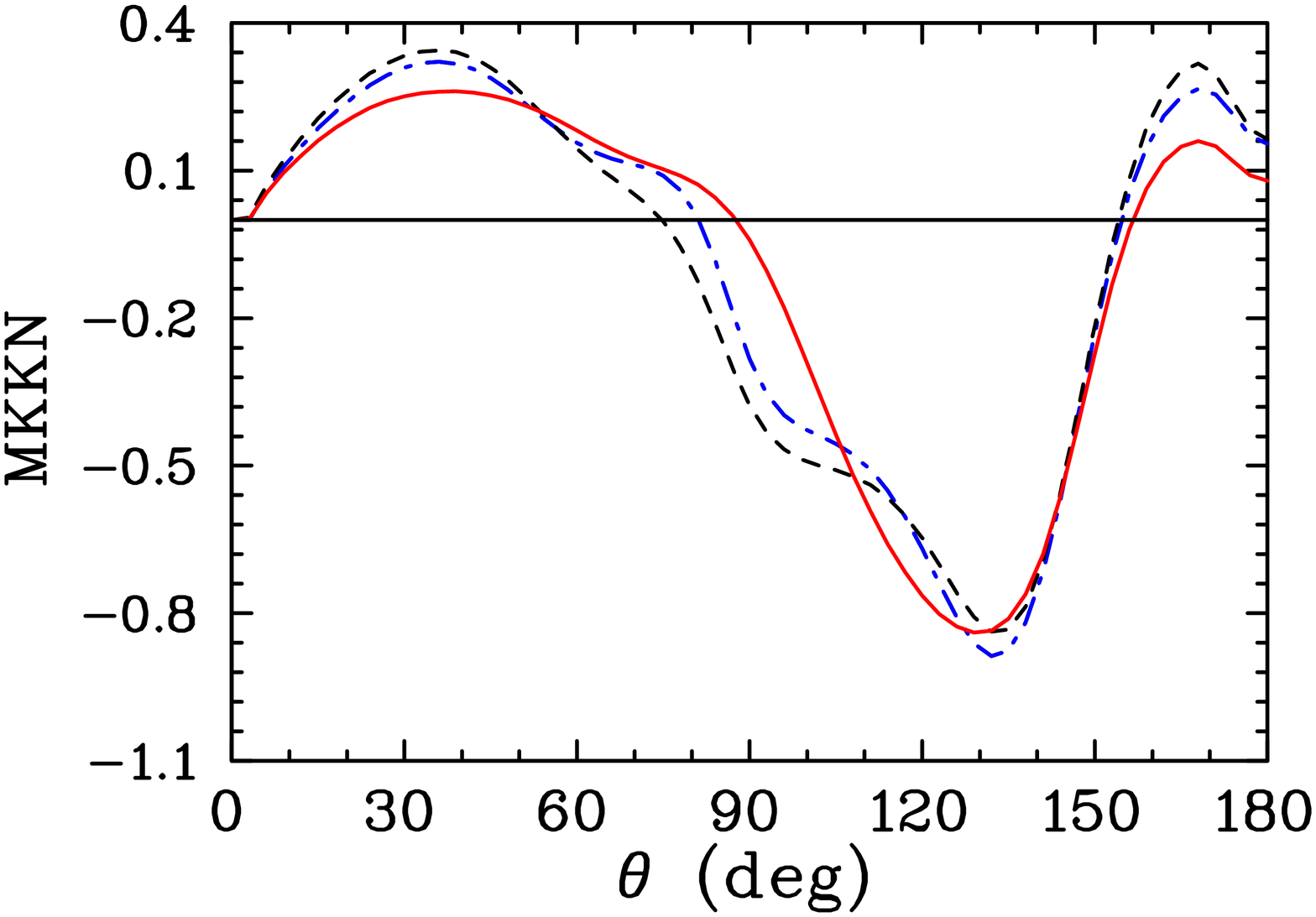}}
\vspace{3mm}
\caption{(Color online) Angular distributions of polarized observables 
	around the COSY resonance~\protect\cite{wasa_prl}: W = 2350~MeV
        (left panels), 2380~MeV (middle panels), and 2410~MeV 
	(right panels). Notation as in Fig.~\protect\ref{fig:g3}. 
	\label{fig:g6}}
\end{figure*}

In Figs.~\ref{fig:g4} -- \ref{fig:g6}, the behavior of many 
other observables is compared to that shown by $A_y$. 
Above 2.4~GeV, the fits are almost unconstrained by data and, 
while the differences are very large, no single measurement 
would allow a reliable PWA. Observables such as MSSN, MKKN, 
MSNK, and NKNS, involve 3 spins and are difficult to measure.
Of these, only MSSN has been measured, and only for $pp$ scattering,
at PSI~\cite{psi,SAID}, with a maximum energy below 600~MeV.
Triple-polarization measurements are extremely difficult and depend 
on the apparatus experimentalists have available. 
A typical case is described by G\"ulmez \textit{et al.}, 
at LAMPF which required the measurement of linear combinations of
observables~\cite{Gulmez}.  
An observable translation guide is given in Table~\ref{tab:conv}. 
The best choice, beyond a single-spin asymmetry~\cite{notation} 
$A_y =  P = P_{n000} = P_{0n00} = A_{00n0} = A_{000n}$ ,
would be the measurement of 2 spins, allowing a test at or below 
the energy of the COSY experiment, where more data are available 
to constrain a fit. 
%%%%%%%%%%%%%%%%%%%%%%%%%%%%%%%%%%%%%%%%%%%%%%%%%%%%%%
\begin{table}[htb!]
\centering
\caption{Sign convention and notation. Bystricky, Lehar, and 
	Winternitz~\protect\cite{le78} give explicit 
	definitions, but some signs differ from 
	SAID~\protect\cite{SAID}. The Bystricky symbols  
	$D$, $K$, $M$, and $N$ denote the depolarization, 
	polarization transfer, and contributions of two 
	initial polarizations to the final polarizations 
	of the of the scattered and recoil particles, 
	respectively.}
\vspace{2mm}
{\begin{tabular}{|c|c|}
\hline
SAID & Bystricky\\
\hline
A   &  $D_{s^\prime 0k0}$\\
RP  &  $D_{k^\prime 0s0}$\\
DT  &  $K_{0nn0}$\\
RPT &  $-K_{0k^{\prime\prime}s0}$\\
\hline
MSNK&  $M_{s^\prime 0nk}$\\
MKKN&  $M_{k^\prime 0kn}$\\
MSSN&  $M_{s^\prime 0sn}$\\
NKNS&  $N_{0k^{\prime\prime}ns}$\\
\hline
\end{tabular}} \label{tab:conv}
\end{table}
%%%%%%%%%%%%%%%%%%%%%%%%%%%%%%%%%%%%%%%%%%%%%%%%%%%%%%

In Figs.~\ref{fig:g4} -- \ref{fig:g6}, the above 
observables are given as angular distributions at three 
energies near to the 2.38~GeV structure. Here, RPT is 
an interesting 
possibility, as the pole and non-pole fits differ 
significantly at CM energies below 2.38~GeV. At angles 
near 75$^\circ$, the discrepancy between the non-pole 
and pole fits is larger than found in a comparison of 
SP07 and the pole fit. A similar effect is seen in DT 
for angles near 115$^\circ$. 

Looking over Figs.~\ref{fig:g3} -- ~\ref{fig:g6}, one can
see that new measurements of $np$ RP, A, DT, and RPT, with 
a precision comparable to previous measurements from LAMPF 
will provide important constraints for the fit.  For 
instance, LAMPF data (which have a limit of 800~MeV) have 
statistical uncertainties of the order $\Delta(RP)\sim0.05$ 
and $\Delta(A)\sim0.05$~\cite{ba85q}, 
$\Delta(DT)\sim0.03$~\cite{mc93}, and 
$\Delta(RPT)\sim0.04$~\cite{mc91}.

%%%%%%%%%%%%%%%%%%%%%%%%%%%%%%%%%%%%%%%%%%%%%%%%%%%%%%
%%%%%%%%%%%%%%%%%%%%%%%%%%%%%%%%%%%%%%%%%%%%%%%%%%%%%%
%%%%%%%%%%%%%%%%%%%%%%%%%%%%%%%%%%%%%%%%%%%%%%%%%%%%%%
%%%%%%%%%%%%%%%%%%%%%%%%%%%%%%%%%%%%%%%%%%%%%%%%%%%%%%

\begin{figure*}[th]
\centerline{
\includegraphics[height=0.35\textwidth, angle=90]{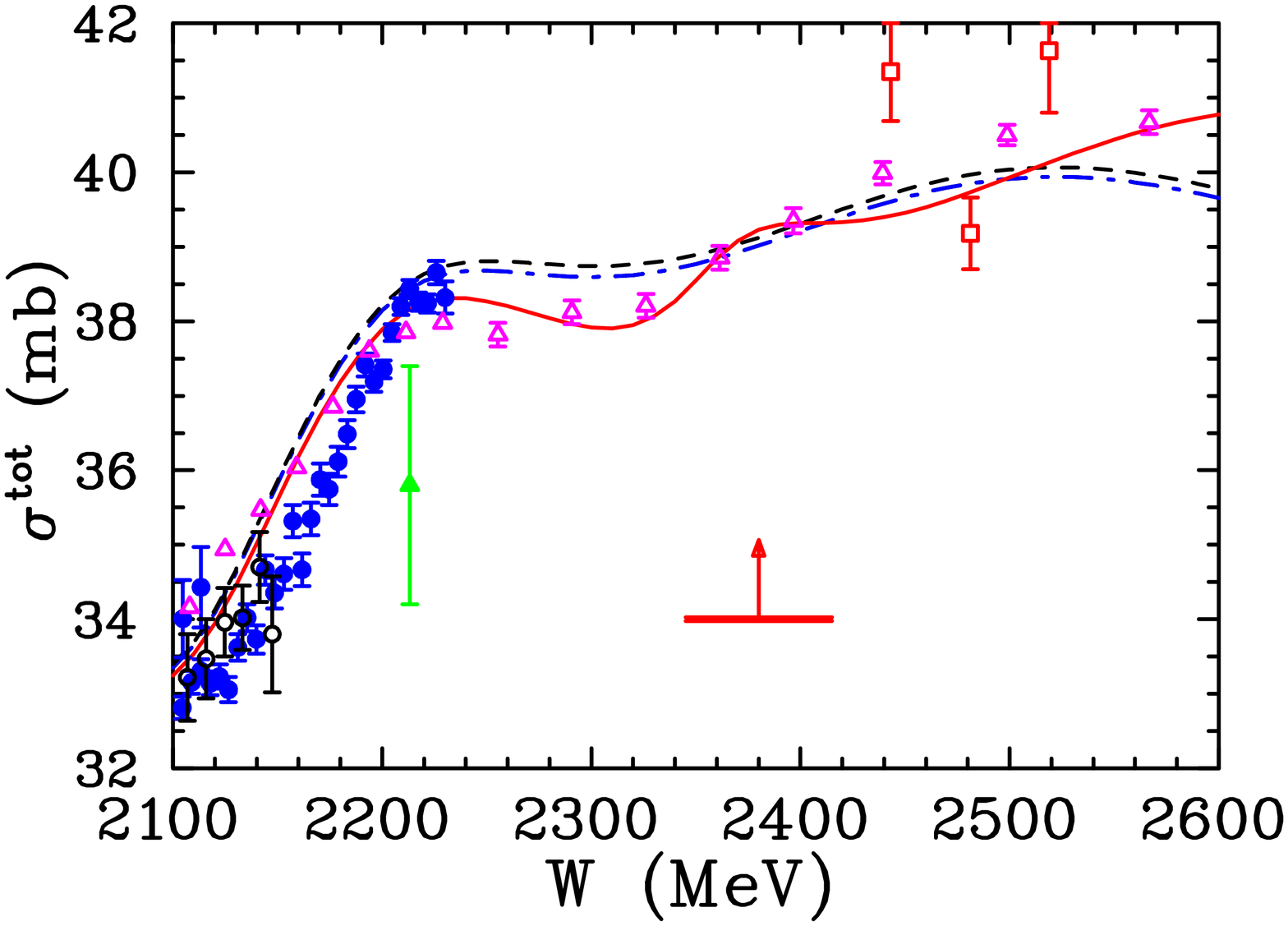}\hfill
\includegraphics[height=0.35\textwidth, angle=90]{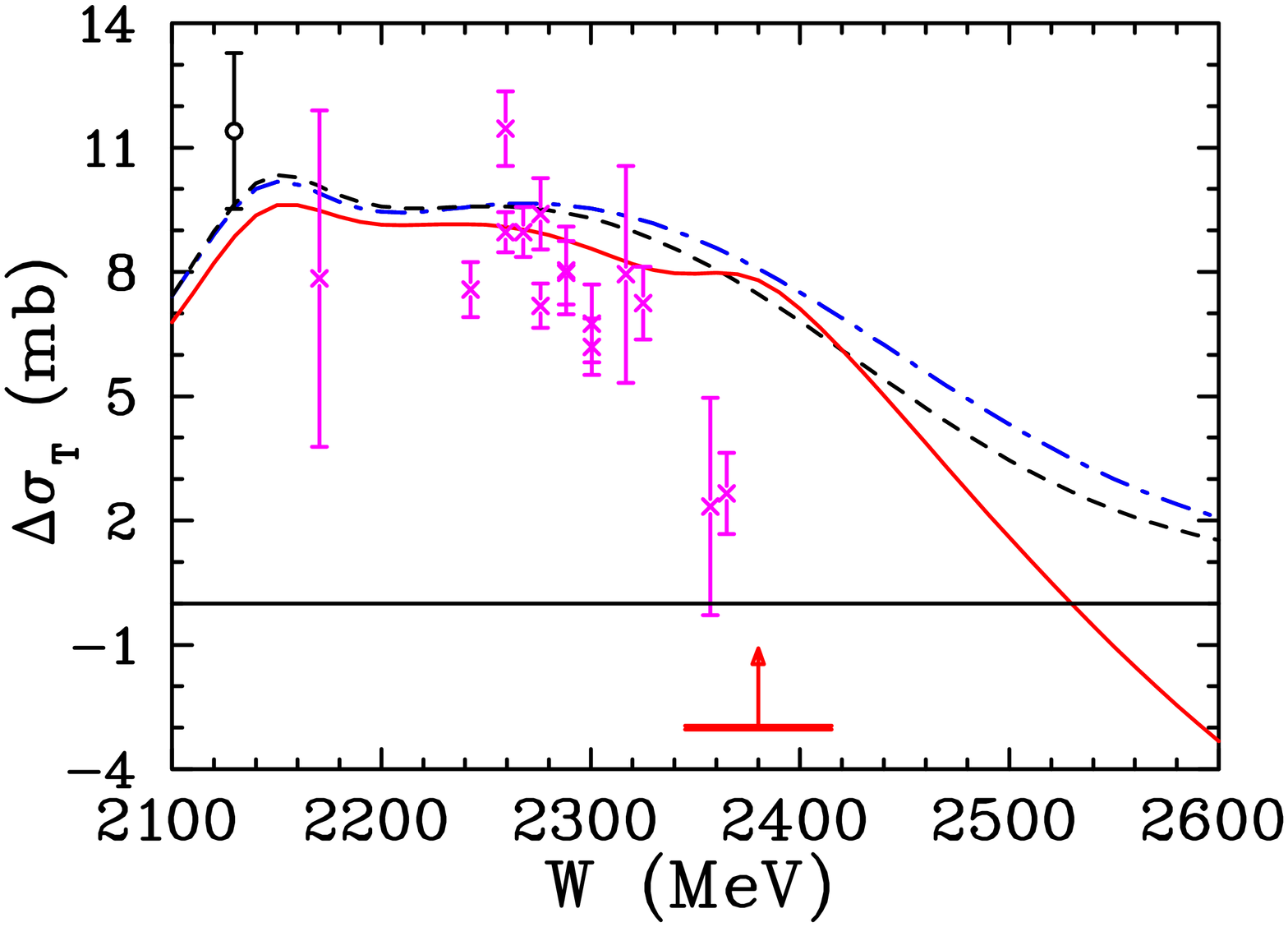}\hfill
\includegraphics[height=0.35\textwidth, angle=90]{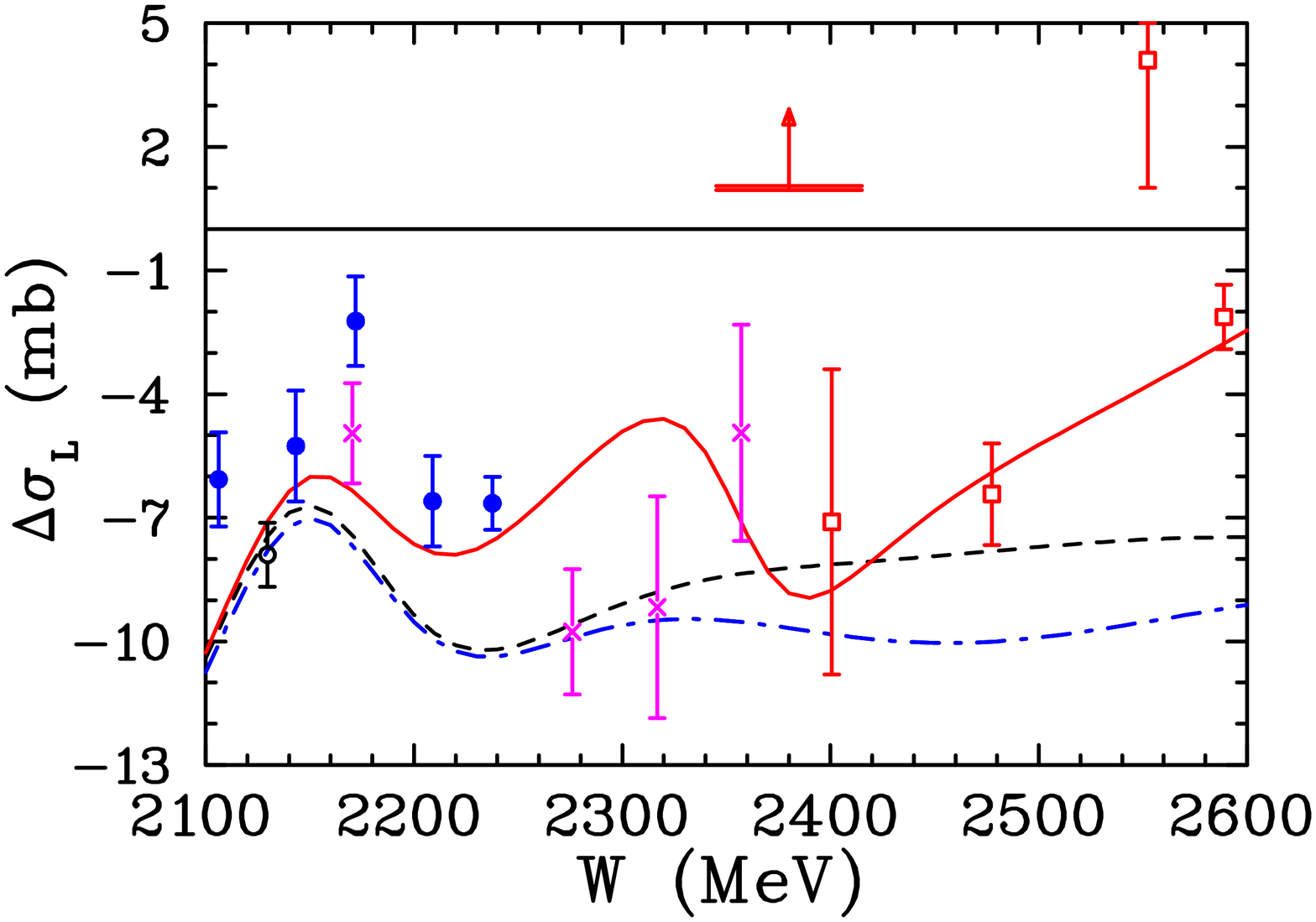}}
\vspace{3mm}
\caption{(Color online) Total cross sections near the COSY 
	resonance~\protect\cite{wasa_prl}: $\sigma^{tot}$
        (left panel) [LAMPF data~\protect\cite{li82,ab01} 
	shown as blue filled circles, PSI data~\protect\cite{gr85} 
	shown as black open circles, PPA data~\protect\cite{de73}
	shown as magenta open triangles, BNL 
	data~\protect\cite{pa64} shown as green filled triangles,
	and JINR data~\protect\cite{sh04} shown as red open sqares], 
	$\Delta\sigma_T$ (middle panel) [PSI data~\protect\cite{bi91}
        shown as black open circles and SACLAY data~\protect\cite{le87,
	fo91,ba94} shown as magenta crosses], and 
        $\Delta\sigma_L$ (right panel) [PSI data~\protect\cite{bi91}
        shown as black open circles, LAMPF data~\protect\cite{be64}
        shown as blue filled circles, Saclay data~\protect\cite{le87} 
	shown as magenta crosses, and JINR 
	data~\protect\cite{sh04,ad96} shown as red open squares]. 
	Notation for SAID curves as in Fig.~\protect\ref{fig:g3}.
        \label{fig:g7}}
\end{figure*}

The total cross section, $\sigma^{\rm tot}$, and spin-dependent neutron-proton total 
cross-section differences, $\Delta\sigma_T$ and $\Delta\sigma_L$
are compared to SAID fits and predictions 
in Fig.~\ref{fig:g7}. The SP07 and pole fits for these quantities were
compared in Ref.~\cite{wasa_prc} but are included here for completeness.
The pole fit differs most from the non-pole fits in $\Delta\sigma_L$, where
existing data is not sufficiently precise to clearly distinguish between these
alternatives. Improved measurements of this quantity, with 
uncertainties comparable to the LAMPF~\cite{be64} measurements, 
of order of $\Delta\sigma_L\sim 10\%$ would greatly improve this comparison.

%%%%%%%%%%%%%%%%%%%%%%%%%%%%%%%%%%%%%%%%%%%%%%%%%%%%%%
%%%%%%%%%%%%%%%%%%%%%%%%%%%%%%%%%%%%%%%%%%%%%%%%%%%%%%

%%%%%%%%%%%%%%%%%%%%%%%%%%%%%%%%%%%%%%%%%%%%%%%%%%%%%%
\section{Summary and Conclusions}
\label{sec:conc}

Motivated by the COSY dibaryon observation, at a CM
energy of 2.38 GeV, in $np$ scattering and two-pion 
production processes, we have made a detailed study 
of possible fits and predictions for $np$ scattering 
observables based on the SAID analysis, with and 
without the contribution of a pole in the $^3D_3$-$^3G_3$ 
coupled waves. 

Given the scarcity of $np$ scattering data above the 
structure seen in $A_y$, the most reliable source of 
information should come from either improved 
measurements of $A_y$ at energies slightly below the
COSY measurement or measurements of two-spin polarization 
quantities showing sizeable deviations between the pole 
and non-pole predictions. Improved measurements of 
$\Delta\sigma_L$ would also be useful.

The precision achieved in previous LAMPF measurements 
should be sufficient to distinguish between the fit
alternatives presented here. For the two-spin observables
of interest, measurements could be confined to 
intermediate angles. The pole fit displays a rapid energy 
variation which would require a fine energy binning and 
measurements spanning the width of the COSY dibaryon. 

%%%%%%%%%%%%%%%%%%%%%%%%%%%%%%%%%%%%%%%%%%%%%%%%%%%%%%
\acknowledgments

We thank M.~Bashkanov, M.~W.~McNaughton, H.M.~Spinka, and 
E.A.~Strokovsky for comments on the feasibility of future 
measurements. This work is supported, in part, by the 
U.S.~Department of Energy, Office of Science, Office of 
Nuclear Physics, under Award Number DE--SC0014133.

%%%%%%%%%%%%%%%%%%%%%%%%%%%%%%%%%%%%%%%%%%%%%%%%%%%%
\clearpage
%%%%%%%%%%%%%%%%%%%%%%%%%%%%%%%%%%%%%%%%%%%%%%%%%%%%%%%%%%%%%%%%%%%

%%%%%%%%%%%%%%%%%%%%%%%%%%%%%%%%%%%%%%%%%%%%%%%%%%%%%%%%%%%%%%%%%

%%%%%%%%%%%%%%%%%%%%%%%%%%%%%%%%%%%%%%%%%%%%%%%%%%%%%

\begin{thebibliography}{0}
\bibitem{notation}
               The $P$ and $A$ multi-index notation is explained in
               Section 3 of Ref.~\cite{le78}.
\end{thebibliography}
\end{document}